\let\includefigures=\iftrue
%
% the following is to use blackboard bold fonts --
\let\useblackboard=\iftrue
%
% activate this if you don't have them.
%\let\useblackboard=\iffalse
%
% You might also need to remove this line.
\newfam\black
\input harvmac.tex
\input epsf
%\draftmode
\includefigures
\message{If you do not have epsf.tex (to include figures),}
\message{change the option at the top of the tex file.}
\input epsf
\def\figin{\epsfcheck\figin}\def\figins{\epsfcheck\figins}
\def\epsfcheck{\ifx\epsfbox\UnDeFiNeD
\message{(NO epsf.tex, FIGURES WILL BE IGNORED)}
\gdef\figin##1{\vskip2in}\gdef\figins##1{\hskip.5in}% blank space instead
\else\message{(FIGURES WILL BE INCLUDED)}%
\gdef\figin##1{##1}\gdef\figins##1{##1}\fi}
\def\DefWarn#1{}
\def\figinsert{\goodbreak\midinsert}
\def\ifig#1#2#3{\DefWarn#1\xdef#1{fig.~\the\figno}
\writedef{#1\leftbracket fig.\noexpand~\the\figno}%
\figinsert\figin{\centerline{#3}}\medskip\centerline{\vbox{\baselineskip12pt
\advance\hsize by -1truein\noindent\footnotefont{\bf Fig.~\the\figno:} #2}}
\bigskip\endinsert\global\advance\figno by1}
%%%
\else
\def\ifig#1#2#3{\xdef#1{fig.~\the\figno}
\writedef{#1\leftbracket fig.\noexpand~\the\figno}%
%\figinsert\figin{\centerline{#3}}\medskip\centerline{\vbox{\baselineskip12pt
%\advance\hsize by -1truein\noindent\footnotefont{\bf Fig.~\the\figno:} #2}}
%\bigskip\endinsert
\global\advance\figno by1}
\fi
\useblackboard
\message{If you do not have msbm (blackboard bold) fonts,}
\message{change the option at the top of the tex file.}
\font\blackboard=msbm10 scaled \magstep1
\font\blackboards=msbm7
\font\blackboardss=msbm5
\textfont\black=\blackboard
\scriptfont\black=\blackboards
\scriptscriptfont\black=\blackboardss

\else

\fi
% *************************************
%\draft
%
\def\yboxit#1#2{\vbox{\hrule height #1 \hbox{\vrule width #1
\vbox{#2}\vrule width #1 }\hrule height #1 }}
\def\fillbox#1{\hbox to #1{\vbox to #1{\vfil}\hfil}}
\def\ybox{{\lower 1.3pt \yboxit{0.4pt}{\fillbox{8pt}}\hskip-0.2pt}}

\def\Coh{{\bf Coh\ }}

\def\mapr{\mathop{\longrightarrow}\limits}
\def\dmapr{\mathop{\Longrightarrow}\limits}
\def\grade{\varphi}
\def\cpxdot{\cdot}

\def\bA{{\bar A}}
\def\bB{{\bar B}}

\def\bZ{{\bar Z}}

\def\p{\partial}
\def\delbar{{\bar\partial}}

\def\half{{1\over 2}}
\def\Tr{{{\rm Tr~ }}}
\def\tr{{\rm tr\ }}

\def\Im{{\rm Im\hskip0.1em}}

\def\lcm{{\rm lcm}}

\def\ket#1{|#1\rangle}

\def\CA{{\cal A}}

\def\CD{{\cal D}}

\def\CF{{\cal F}}

\def\CT{{\cal T}}

\def\CM{{\cal M}}
\def\CN{{\cal N}}
\def\CO{{\cal O}}
\def\CP{{\cal P}}

\def\P{\BP}

\def\II{\relax{I\kern-.10em I}}

\def\IIb{{\II}b}

\def\MK{{\CM_k}}

\def\IZ{\relax\ifmmode\mathchoice
{\hbox{\cmss Z\kern-.4em Z}}{\hbox{\cmss Z\kern-.4em Z}}
{\lower.9pt\hbox{\cmsss Z\kern-.4em Z}}
{\lower1.2pt\hbox{\cmsss Z\kern-.4em Z}}\else{\cmss Z\kern-.4em
Z}\fi}
\def\IB{\relax{\rm I\kern-.18em B}}
\def\IC{{\relax\hbox{$\inbar\kern-.3em{\rm C}$}}}
\def\ID{\relax{\rm I\kern-.18em D}}
\def\IE{\relax{\rm I\kern-.18em E}}
\def\IF{\relax{\rm I\kern-.18em F}}
\def\IG{\relax\hbox{$\inbar\kern-.3em{\rm G}$}}
\def\IGa{\relax\hbox{${\rm I}\kern-.18em\Gamma$}}
\def\IH{\relax{\rm I\kern-.18em H}}
\def\IK{\relax{\rm I\kern-.18em K}}
\def\IN{\relax{\rm I\kern-.18em N}}
\def\IP{\relax{\rm I\kern-.18em P}}
%\def\IX{\relax{\rm X\kern-.01em X}}
%this doesn't work

%
\def\inbar{\,\vrule height1.5ex width.4pt depth0pt}
\def\mod{{\rm\; mod\;}}

\def\p{\partial}

\font\cmss=cmss10 \font\cmsss=cmss10 at 7pt
\def\IR{\relax{\rm I\kern-.18em R}}

\def\Hom{{\rm Hom}}
\def\Ext{{\rm Ext}}

\def\BZ{Z} % for now
\def\BP{\IP}
\def\BR{\IR}
\def\BC{\IC}

\def\lp10{l_P^{10}}
\def\lp11{l_P^{11}}
\def\R11{R_{11}}

\Title{\vbox{\baselineskip12pt\hbox{hep-th/0011017}
\hbox{RUNHETC-2000-42}}}
{\vbox{
\centerline{D-branes, Categories}
\smallskip
\centerline{and $\CN=1$ Supersymmetry}}}
\smallskip
\centerline{Michael R. Douglas\footnote{$^\&$}{
Louis Michel Professor}}
\medskip
\centerline{Department of Physics and Astronomy}
\centerline{Rutgers University }
\centerline{Piscataway, NJ 08855--0849}
\smallskip
\centerline{\it and}
\smallskip
\centerline{I.H.E.S., Le Bois-Marie, Bures-sur-Yvette, 91440 France}
\medskip
\centerline{\tt mrd@physics.rutgers.edu}
\bigskip
\noindent
We show that boundary conditions in
topological open string theory on Calabi-Yau manifolds
are objects in the derived category of coherent sheaves,
as foreseen in the homological mirror symmetry proposal of Kontsevich.
Together with conformal field theory considerations,
this leads to a precise criterion determining the BPS branes
at any point in CY moduli space, completing the proposal of $\Pi$-stability.

\Date{October 2000}
\nref\aspdon{P. Aspinwall and R. Donagi,
``The Heterotic String, the Tangent Bundle, and Derived Categories,''
hep-th/9806094.}
\nref\beilinson{A. A. Beilinson, ``Coherent sheaves on $\IP^n$ and
  problems of linear algebra'', {\it Funct. Anal. Appl.}
  {\bf 12} (1978) 214--216.}
\nref\bbd{A. Beilinson, J. Bernstein and P. Deligne,
``Faiseaux pervers,'' Ast\'erisque 100 (1982).}
\nref\bdl{M. Berkooz, R. Leigh and M. R. Douglas,
``Branes Intersecting at Angles,'' Nucl.Phys. B480 (1996) 265-278,
hep-th/9606139.}
\nref\bsv{M. Bershadsky, V. Sadov, and C. Vafa,
``D-Branes and Topological Field Theories,''
Nucl.Phys. B463 (1996) 420-434, hep-th/9511222.}
\nref\boflop{A. Bondal and D. Orlov, ``Semiorthogonal decomposition
for algebraic varieties,'' math.AG/9506012.}
\nref\bondalorlov{A. Bondal and D. Orlov, ``Reconstruction of a variety
from the derived category and groups of autoequivalences,''
alg-geom/9712029.}
%\nref\bradlow{Bradlow}
\nref\bridgeland{T. Bridgeland, ``Flops and derived categories,''
math.AG/0009053.}
\nref\bkr{T. Bridgeland, A. King, and M. Reid, ``Mukai implies McKay'', 
  math.AG/9908027.}
\nref\bdlr{I. Brunner, M. R. Douglas, A. Lawrence, and C. R\"omelsberger,
  ``D-branes on the quintic,'' hep-th/9906200.}
\nref\candelas{P. Candelas, X. C. de la Ossa, P. S. Green, and 
  L. Parkes, ``A pair of
  Calabi--Yau manifolds as an exactly soluble superconformal theory'',
  Nucl. Phys. {\bf B359} (1991) 21.}
\nref\dlp{J. Dai, R. Leigh and J. Polchinski,
Mod.Phys.Lett.A4:2073-2083,1989.}
\nref\denev{F. Denef, ``Supergravity flows and D-brane stability,''
JHEP 0008 (2000) 050, hep-th/0005049.}
\nref\dd{D.-E. Diaconescu and M. R. Douglas, ``D-branes on Stringy
Calabi-Yau Manifolds,'' hep-th/0006224.}
\nref\ddtwo{D.-E. Diaconescu and M. R. Douglas, work in progress.}
\nref\dg{D.-E. Diaconescu and J. Gomis, ``Fractional branes and boundary
  states in orbifold theories,'' hep-th/9906242.}
\nref\dtopics{M. R. Douglas, ``Topics in D-geometry'',
  {\it Class. Quant. Grav.} {\bf 17} (2000) 1057, hep-th/9910170.}
\nref\pistable{M. R. Douglas, B. Fiol, and C. R\"omelsberger,
  ``Stability and BPS branes'', hep-th/0002037.}
\nref\noncompact{M. R. Douglas, B. Fiol, and C. R\"omelsberger, 
  ``The spectrum of BPS branes on a noncompact Calabi--Yau'', 
  hep-th/0003263.}
\nref\mike{M. R. Douglas, lecture at Cern, July 2000.}
\nref\fiolmarino{B. Fiol and M. Marino,
``BPS states and algebras from quivers,''
JHEP 0007 (2000) 031, hep-th/0006189.}
\nref\gm{S. I. Gelfand and Y. Manin,
  ``Homological Algebra,'' Springer 1991.}
\nref\gepner{D.~Gepner, ``Space-Time Supersymmetry In Compactified
String Theory And Superconformal Models," Nucl. Phys. {\bf B296}, 757 (1988).}
\nref\govj{S. Govindarajan and T. Jayaraman, 
  ``On the Landau--Ginzburg description
  of boundary CFTs and special Lagrangian submanifolds'', 
  hep-th/0003242.}
\nref\govjs{S. Govindarajan, T. Jayaraman, and T. Sarkar, 
  ``World sheet approaches to D-branes 
  on supersymmetric cycles'', hep-th/9907131.}
\nref\govgep{S. Govindarajan and T. Jayaraman,  
``D-branes, Exceptional Sheaves and Quivers on 
Calabi-Yau manifolds: From Mukai to McKay,'' hep-th/0010196.}
\nref\greene{B. R. Greene and Y. Kanter, 
``Small Volumes in Compactified String Theory,''
Nucl.Phys. B497 (1997) 127-145, hep-th/9612181.}
\nref\hm{J. Harvey and G. Moore, ``On the algebras of BPS states,''
Comm. Math. Phys. 197 (1998) 489-519, hep-th/9609017.}
\nref\hiv{K. Hori, A. Iqbal, and C. Vafa, 
  ``D-branes and mirror symmetry'', hep-th/0005247.}
\nref\horja{R.~P.~Horja, ``Hypergeometric functions and mirror symmetry 
in toric varieties'', math.AG/9912109}
\nref\itonak{Y. Ito and H. Nakajima, ``McKay correspondence and
  Hilbert schemes in dimension three'', math.AG/9803120.}
\nref\joyce{D.~Joyce, ``On counting special Lagrangian homology
3-spheres,'' hep-th/9907013.}
\nref\kamg{S.~Kachru and J.~McGreevy, ``Supersymmetric three-cycles
and (super)symmetry breaking,'' hep-th/9908135.}
\nref\kachmirror{S. Kachru, S. Katz, A. Lawrence and J. McGreevy,
``Open string instantons and superpotentials,''
Phys. Rev. D62 (2000) 026001, hep-th/9912151 and
``Mirror symmetry for open strings,'' hep-th/0006047.}
\nref\kaporlov{A. Kapustin and D. Orlov, ``Vertex Algebras, 
Mirror Symmetry, And D-Branes: The Case Of Complex Tori,''
hep-th/0010293.}
\nref\lerche{P. Kaste, W. Lerche, C. A. L\"utken, and J. Walcher,
  ``D-branes on K3 fibrations,'' hep-th/9912147.}
\nref\keller{B. Keller, ``Introduction to A-infinity Algebras
and Modules,'' math.RA/9910179.}
\nref\kontsevich{M.~Kontsevich, ``Homological algebra of mirror 
  symmetry'', alg-geom/9411018.}
\nref\mayrgep{P. Mayr, ``Phases of Supersymmetric D-branes on 
Kaehler Manifolds and the McKay correspondence,'' hep-th/0010223.}
\nref\merkulov{S. Merkulov, ``Strong homotopy algebras of
a Kahler manifold,'' math.AG/9809172.}
\nref\ooy{H. Ooguri, Y. Oz and Z. Yin,
``D-Branes on Calabi-Yau Spaces and Their Mirrors,''
Nucl.Phys. B477 (1996) 407-430, hep-th/9606112.}
\nref\oz{Y. Oz, T. Pantev and D. Waldram,
``Brane-Antibrane Systems on Calabi-Yau Spaces,'' hep-th/0009112.}
\nref\pol{J. Polchinski, ``TASI Lectures on D-branes,'' hep-th/9611050.}
\nref\polishchuk{A. Polishchuk, ``Homological Mirror Symmetry with
Higher Products,'' math.AG/9901025.}
\nref\rs{A. Recknagel and V. Schomerus, ``D-branes in Gepner models'',
  Nucl. Phys. {\bf B531} (1998) 185, hep-th/9712186.}
\nref\reid{M. Reid, ``McKay correspondence'', alg-geom/9702016.}
\nref\reidbourbaki{M. Reid, ``La correspondance de McKay'', S\'eminaire
  Bourbaki (novembre 1999), no. 867, math.AG/9911165.}
\nref\segal{G. Segal, unpublished.}
\nref\seidelthomas{P. Seidel and R. Thomas, ``Braid group actions on
derived categories of sheaves,'' math.AG/0001043.}
\nref\sharpe{E.~Sharpe, ``Kaehler cone substructure,''
Adv.\ Theor.\ Math.\ Phys.\  {\bf 2}, 1441 (1999) [hep-th/9810064].}
\nref\sharpedc{E. Sharpe, 
``D-Branes, Derived Categories, and Grothendieck Groups,''
hep-th/9902116.}
\nref\taylor{W. Taylor IV, ``D-brane field theory on compact spaces,''
Phys.Lett. B394 (1997) 283-287, hep-th/9611042.}
\nref\thomas{R. P. Thomas, ``Mirror symmetry and actions of braid
groups on derived categories,'' math.AG/0001044.}
\nref\thomasdc{R. P. Thomas, ``Derived categories for the working
mathematician,'' math.AG/0001045.}
\nref\tomasiello{A. Tomasiello,
``Projective resolutions of coherent sheaves 
and descent relations between branes,'' hep-th/9908009.}
\nref\tomgep{A. Tomasiello, ``D-branes on Calabi-Yau manifolds and helices,''
hep-th/0010217.}
\nref\vafa{C. Vafa, ``Extending Mirror Conjecture 
to Calabi-Yau with Bundles,'' hep-th/9804131.}
\nref\wittentop{E. Witten, ``Chern-Simons Gauge Theory as a String
Theory,'' hep-th/9207094.}
\nref\witten{E. Witten, ``Phases of $N{=}2$ theories in two dimensions,''
  {\it Nucl. Phys.} {\bf B403} (1993) 159, hep-th/9301042.}
\nref\wittensmall{E. Witten, ``Small Instantons in String Theory,''
Nucl. Phys. B460 (1996) 541-559; hep-th/9511030.}
\nref\wittenK{E. Witten, ``D-Branes And K-Theory,''
JHEP 9812 (1998) 019, hep-th/9810188.}
%
% forward equation references
%
\newsec{Introduction}

Over the last year, the basic elements of a picture of BPS D-branes in
weakly coupled type \II\ string theory on general Calabi-Yau (CY)
backgrounds have been developed, following the lines described in
\dtopics.  Such branes have world-volume theories with $\CN=1$, $d=4$
supersymmetry or its equivalent and by analogy with the study of
supersymmetric field theory, one might expect to be able to get a good
understanding of the observables determined by holomorphic or
``protected'' quantities such as the superpotential and D-flatness
conditions; these are the spectrum of BPS branes and their moduli
spaces of supersymmetric vacua.  
By analogy with the study of $\CN=2$ compactifications using mirror
symmetry, one might even hope to get this understanding and determine
the BPS spectrum everywhere in moduli space (i.e. for string scale
CY's) from a suitable reinterpretation of large volume results.  

In this work, we give a proposal for how to do this, building on the
$\Pi$-stability proposal of \refs{\pistable,\noncompact}, by combining
physical input, notably the theory of boundary conditions in $(2,2)$
CFT and its topological twistings, with the wealth of relevant
mathematics, especially Kontsevich's homological mirror symmetry
proposal \kontsevich\ and the formalism of the derived category.

A longer work with more introductory discussion is in preparation;
here we try to give a relatively concise discussion of the ideas and
results.  In particular, we will not give very precise explanations of
the fairly lengthy mathematical background (mostly homological
algebra), instead focusing on its physical interpretation.  This
background can be found in the standard reference \gm; we also
recommend \thomasdc\ for a nice introduction to the derived category.

A good way to identify observables which can be computed in the large
volume limit is to consider the topologically twisted world-sheet
theory.  We will consider sigma models with CY target.
These models have two twisted versions, the A twisted model whose
observables depend only on (the stringy generalization of) the CY
K\"ahler moduli, and the B twisted model whose observables depend only
on the CY complex moduli.  Stringy corrections are absent in the B
model and thus these observables are computable at large volume;
mirror symmetry can then be used to compute A model observables.

For the closed string, the basic ``topological'' observable is the
prepotential of the $\CN=2$ supergravity obtained by type \II\
compactification on the CY.  Equivalent information is the metric on
moduli space, or the central charges of BPS states at a general point
in moduli space.  In the B model these central charges are the periods
of the holomorphic $3$-form and are computable.

For the open string, the basic topological observables are the set of
allowed topological boundary conditions (which we could call
``topological D-branes''), and the spectrum of open strings between
any pair of topological D-branes.  Physically, these correspond to
D-brane configurations which solve the F-flatness conditions, and the
massless fermionic open strings between any pair of D-branes.
Mathematically, this data can be thought of as defining a category of
brane configurations, in which the objects are topological D-branes,
and the morphisms are open strings.

In principle, this allows computing these observables in the B model
at large volume, and extrapolating them to general points in K\"ahler
moduli space.  However, this is subtle -- the naive extrapolation
according to which B branes are holomorphic bundles or coherent
sheaves at every point in K\"ahler moduli space, is incorrect
\noncompact.

The source of this contradiction is the following: if one makes a large
variation of the K\"ahler moduli, in general a pair of branes which
has aligned BPS central charges (i.e. with the same phase, so
preserving the same $\CN=1$ supersymmetry), can vary into a pair with
arbitrarily related and even anti-aligned central charges, as would be
the case for a brane and an antibrane.  Thus any candidate definition
of ``topological D-brane'' which could make sense throughout K\"ahler
moduli space must be able to describe branes and antibranes on the
same footing, and thus include more objects than coherent sheaves.

Now there are already natural mathematical candidates for the category
of topological D-branes -- in the B model, the derived category of
coherent sheaves, and in the A model the Fukaya category, as proposed
some time ago by Kontsevich \kontsevich.
A lot of evidence has accumulated that this is correct, most notably
in results of Seidel and Thomas \seidelthomas\ and Horja \horja\ which we
discuss below.
Physicists have also suggested various roles for this category
\refs{\aspdon,\oz,\sharpedc,\tomasiello}.\footnote*{
After the completion of this work, the interesting work 
\kaporlov\ appeared, which discusses branes on tori 
in terms of the derived category.}

In the present work, we give physical arguments that this is correct.
These follow the spirit of Witten's argument that the topological
class of a D-brane on the space $X$ is a K theory class on $X$
\wittenK: we
define objects to be bound states of D-branes modulo a relation which
equates two configurations which differ by adding cancelling
brane-antibrane pairs.  However we will keep track of far more
information along the way -- essentially, all of the morphisms between
objects -- and thus the result, the derived category, makes much finer
distinctions than K theory.  As an example, whereas the K theory class
of a D$0$-brane does not depend on what point it sits at, in the
derived category every point is a distinct object.

Technically, this begins with the discussion of topological open
string theory as in \wittentop, but we then generalize the BRST
operator to carry additional ``homological'' information associated
with the Chan-Paton factors.  This will allow treating complexes of
boundary states as objects; we will then physically motivate imposing
equivalence relations under adding brane-antibrane pairs, under
$Q$-exact variations of the additional data (homotopy equivalence),
and under complex gauge transformations, and argue that the result is
the derived category formed from the original category of boundary
conditions.

Having some understanding of the topological D-branes, we will then
discuss their relation to physical D-branes.  The primary result in this
direction is a flow of the gradings of objects and morphisms under
variation of K\"ahler moduli, for which we give a simple CFT argument
as well as explicit examples.  

This result allows predicting the masses of bosons in chiral
multiplets at arbitrary points in K\"ahler moduli space, and is thus
the key to determining bound state formation and stability.  In
particular, there is an inconsistency in the CFT interpretation of
morphisms of negative degree -- they would correspond to operators
of negative dimension -- which forces certain topological D-branes
to drop out of the physical spectrum.  This gives a direct CFT argument
for the $\Pi$-stability condition proposed in \pistable.

This agreement suggests that we look for a reformulation of
$\Pi$-stability which does not require an abelian category.
The direct generalization of subobjects and exact sequences to the
derived category is the ``distinguished triangle,'' and we discuss
this notion and show how it unifies different versions of
the bound state and decay processes involving branes and antibranes.

Combining this with the previous results leads to a proposal for a
reformulated stability condition, which although similar to 
$\Pi$-stability does not require a preexisting notion of abelian
category.  This appears to be a good candidate for a mathematically
precise definition of a BPS brane on Calabi-Yau manifold, and is
quite concrete in simple examples, predicting marginal stability lines
and new BPS branes.

\newsec{BPS D-branes, grading, and images}

Our starting point will be a $(2,2)$ SCFT such as a sigma model on a
Calabi-Yau threefold or a Gepner model, with integrally quantized
$U(1)$ charges, so that it can be used to define a type \II\ string
compactification with $d=4$, $\CN=2$ supersymmetry.

The most basic attributes of a BPS D-brane $B$ (we take it to be a
particle in four dimensions for this discussion) are its RR charge
$Q(B)$ and its BPS central charge $Z(Q(B))$ or $Z(B)$.  These are
discussed in many references such as \greene.  We will consider B-type
branes, for which $Q(B)$ is essentially the K theory class.  $Z(B)$
depends on $Q(B)$ and on a point in the stringy K\"ahler moduli space
$\MK$ (best defined as the complex structure moduli space of the
mirror CY).

A quantity which will be particularly
important for us is the phase of the BPS central charge.  As in
\pistable, we define the {\it grade} $\grade$ or $\grade(B)$ 
of a brane to be this phase
normalized so that branes and antibranes have 
$\grade(B)-\grade(\bB)=1 \mod 2$:
\eqn\defgrade{
\grade(B) = {1\over\pi} \Im \log Z(B) .
}
We will need to extend this from the circle $[0,2)$ to a real number;
thus there is a $2\BZ$ ambiguity to be fixed for each brane.
If we do this at some point in moduli space,
the grades can be defined elsewhere by analytic continuation
of $Z$; we will justify this shortly.

At large volume, BPS branes are either A branes (special Lagrangian
manifolds carrying flat connections) or B branes (coherent sheaves
carrying Hermitian Yang-Mills connections).  The most basic
observables in the classical theory are the massless fermion spectrum
between a pair of branes.  For B branes which fill the CY, these can
be obtained by the standard arguments of Kaluza-Klein reduction:
the oriented massless fermionic strings from branes carrying the
bundle $E$ to branes carrying the bundle $F$ are elements of the
complex cohomology group $H^{0,q}(M,E^*\otimes F)$.
As we will review, these are the states of the B twisted topological
open string theory, so this result does not obtain stringy
corrections.  The mirror A statement must involve stringy effects as
is discussed in \refs{\wittentop,\kachmirror,\vafa}; 
because of this we prefer the B picture.

\subsec{BPS branes as boundary conditions}

In world-sheet terms, a BPS brane corresponds to a boundary condition
which preserves an $\CN=2$ world-sheet supersymmetry,
as discussed in \ooy.

For A boundary conditions we have (in the open string channel)
$T_L=T_R$, $G^\pm_L=G^\mp_R$, $J_L=- J_R$, and
\eqn\specbc{
e^{i{Q_0\over 2}\phi_L} = 
e^{i\pi\grade} e^{-i{Q_0\over 2}\phi_R}
}
where $J_L=iQ_0\p\phi_L$ and
$J_R=iQ_0{\bar\p}\phi_R$ define the
bosonization of the $U(1)$ current in the $(2,2)$ algebra,
$Q_0=\sqrt{\hat c}$
and $\hat c=3$ for a CY sigma model is the complex dimension
of the CY.
B boundary conditions are similar with $J_R\rightarrow -J_R$
and $\phi_R\rightarrow-\phi_R$.

The operators in \specbc\
are the $\CN=2$ spectral flow operators which directly enter the
space-time supercharge and thus $e^{i\pi \grade}$ directly specifies an
unbroken $\CN=1$ algebra.
In sigma model compactification we have 
$$
e^{iQ_0\phi_L} = \Omega_{ijk} \psi^i\psi^j\psi^k
$$
in terms of the world-sheet fermions and so for A boundary conditions
$\grade$ is precisely as defined in \defgrade.
This is also true for B boundary conditions, of course.
If the two boundary conditions have $\grade_1-\grade_2\in 2\BZ$, there is an
overall unbroken $\CN=1$ space-time supersymmetry; otherwise not.

From \specbc\ we also see that A and B boundary
conditions are Dirichlet and Neumann boundary conditions respectively
on the boson $\phi$, so $\grade$ is analogous to a ``position'' for A
boundary conditions and a ``position on the T-dual circle'' for B
boundary conditions.  We will always use the Dirichlet language in which
$\grade$ is a position and $U(1)$ charge is a ``winding number,'' even
though we discuss B branes, just for pictorial convenience.

The massless Ramond sector can be obtained by spectral flow from
$\CN=2$ primary chiral states, whose $U(1)$ charge $q$ is equal to the
grading $H^{0,q}$ of cohomology in the large volume B brane
discussion.  In this limit the grade $\grade$ of all CY-filling branes
is the same, and there is an $\CN=1$ supersymmetry.  
Spectral flow can be used even
when space-time supersymmetry is broken, but let us postpone 
discussion of this point.

\subsec{Grading as a real number}

We can now explain the extension of $\grade$ from
a number defined modulo $2$ to a real-valued number.
The point is that this extension makes no difference for a
single brane, but as soon as we consider pairs of branes a relative
integer shift of $\grade$ in the boundary condition for one of them will 
modify the spectrum of strings stretched between them.

There is a familiar construction of D-brane world-volume
theories on $S^1$ (or a torus) which is quite analogous. 
\refs{\dlp,\taylor}\ %
Let us take the circumference to be $2$ (like our variable $\grade$);
the idea is simply to define D-branes on $S^1\cong\BR/2\BZ$ as the theory
of an infinite set of image D-branes in $\BR$, located at lattice
points $2 n + \grade$ for $n\in\BZ$,
and quotient by a simultaneous shift in space-time and gauge
transformation.  
As is well known,
if we start with D$p$-branes with world-volume $U(N)$
$p+1$-dimensional super Yang-Mills theory, the resulting theory is
equivalent to $p+2$-dimensional SYM theory on a circle with the T-dual
radius.

In this context, it is clear that the integer part of $\grade$ is not
a physical quantum number but rather is a gauge degree of freedom.  
However, open strings, and other quantities relating 
to a pair of branes, are labelled by an integer,
the difference $m-n$ between the positions of the two images.
On this point, the analogy with the grade is quite precise.

A difference between the two problems is that the periodicity in
the string spectrum is different: for a single D-brane on $S^1$, the
spectrum has the same periodicity as the images, while in $\CN=2$
SCFT there is a more complicated correlation between the $U(1)$ sector
and the rest of the theory, as discussed in \refs{\gepner,\rs}.
The open string partition function factorizes as a sum of products of
partition functions in the two factors, say for strings between branes
$E$ and $F$
$$
Z^{(E,F)} = \sum_p \chi_p^{U(1)} \chi^{(E,F)_{rest}}_p ,
$$
where the structure of the
second factor determines the allowed winding numbers.

The spectrum is periodic under the action of the spectral flow
operator.  A periodicity between space-time bosons is obtained by
acting with its square $e^{iQ_0\phi}$, which shifts the grade by $\hat
c$.  In the case at hand of $\hat c=3$, the true periodicity is $2\hat
c$, as the GSO projection reverses sign under odd shifts.

All this does not invalidate the picture of
a brane as having integer spaced images, but requires us to distinguish
the images.  This gives us a convenient way to picture 
the winding strings corresponding to higher morphisms, which will
be reproduced later in the mathematics.

The antibrane of a brane is defined by reversing the open
string GSO projection.  This can be accomplished by shifting the
fermion number by an odd integer, so these images should be interpreted
as the antibrane and its images.  We will make this precise by requiring
the formalism to be invariant under a simultaneous shift of all gradings
by $1$ and reversal of all K theory classes, and interpreting this
as a gauge symmetry.

\subsec{Topologically twisted open string theory}

A ``category'' of branes will be considered to be the set of
branes themselves (the ``objects'' of the category) and
the spectrum of massless fermionic strings between pairs of branes,
or ``morphisms'' of the category.  This data for BPS branes 
obeys the mathematical axioms of a category -- in particular,
the morphisms have an associative multiplication law which in CFT
terms is essentially the multiplication law in the open string chiral ring.

By extension, one can consider all of the holomorphic data involving
the branes -- the space-time vector and chiral multiplets which
contain these fermions, and the superpotential, to be part of the
``category'' as well.  All of this data has been claimed to be
independent of the K\"ahler class of the CY (for B branes), implicitly
in work on topological open string theory and in Kontsevich's
homological mirror symmetry proposal, and explicitly in the work
\bdlr.  One can also give a simple space-time argument for this \mike,
based on the fact that in type \IIb\ string compactification (in which
the branes can be taken to fill $3+1$ Minkowski dimensions) the
K\"ahler moduli are paired under $\CN=1$ supersymmetry with RR
potentials, which do not have nonderivative couplings in perturbative
string theory.  (This answers the question of what determines the
coordinates on moduli space for which decoupling holds.)

All this should allow determining the category from
large volume considerations.
To better understand this, we now consider the B twisted topological
sigma model with CY target, following \wittentop.  We start with
bosonic coordinates $Z^i$ and $\bZ^{\bar i}$ and their fermionic partners.
After twisting,
the fermions split into 
left and right moving scalars $\eta^{\bar i}$ and $\theta_i$,
and one-forms $\rho^i$.
The simplest boundary conditions are Neumann;\footnote*{ 
One could also use partly Dirichlet boundary
conditions associated to holomorphic submanifolds, but we will instead
get these at a later stage.}
they set
$\p_n Z^i=\p_n \bZ^i$ and
$\theta=(*\rho)=0$.  
One can also couple to holomorphic
bundles, adding a term 
$\Tr P \int Z^*(A) - i \rho^i F_{i\bar j} \eta^{\bar j}$
on the boundary.

States in a hamiltonian quantization of this theory are determined
by their dependence on the zero modes of $Z$ and $\eta^{\bar i}$, and
can thus be regarded as holomorphic $p$-forms.
The BRST operator $Q$ then satisfies $[Q,Z] = 0$, 
$[Q,\bar Z^{\bar i}]=-\eta^{\bar i}$ and $\{Q,\eta\}=0$, which
means that it can be interpreted as a $\delbar$ operator
coupled to $E^* \otimes F$.
Thus, the
topological open string Hilbert space with boundary conditions $(E,F)$
is the $Q$-cohomology, which is $H^*(X, E^* \otimes F)$.
The multiplication law is of course wedge product of forms.

In early studies of D-branes it was found that certain
point-like singularities are allowed and are non-singular
in string theory  \wittensmall.  This motivates allowing more general
coherent sheaves as boundary conditions \hm.
The entire discussion can be generalized to this case
at least formally by replacing $H^p(X, E^* \otimes F)$ with
$\Ext^p(E,F)$, which for a pair of holomorphic bundles is equivalent.
We will explain and justify this point below;
for this reason we switch to use the $\Ext$ notation instead of
cohomology.  We also remind the reader that $\Hom(E,F)\equiv\Ext^0(E,F)$.

The basic topological correlation function is a disk amplitude,
non-zero for a combination of states whose charge adds up to $\hat c=3$.
This is determined by the algebra structure and the integral on
$\Ext^3(E,E)$; this integral can also be regarded as a trace and this
structure defines a ``Frobenius category.''  We can also use the integral
to define Serre duality, which here identifies 
$\Ext^{3-i}(E,F)\cong \Ext^i(F,E)^*$. 

It is known \refs{\kontsevich,\segal}\ that in general topological open
string theory can correspond to an $A_\infty$ category \keller, as
appeared in Kontsevich's original proposal.  An explicit construction
of an $A_\infty$ structure on the category of coherent sheaves appears
in \refs{\merkulov,\polishchuk}; the higher products are essentially
correlation functions in holomorphic Chern-Simons theory
(the third order product was already discussed in \wittentop).
They are related to the Massey products, which encode
the obstruction theory or equivalently the physical superpotential.
This structure is very useful in studying deformations, as we will
discuss in subsequent work, but we will not need it for our present
considerations.  In particular, the $A_\infty$ structure defined in
\polishchuk\ is a
``minimal model,'' i.e. an $A_\infty$ category with $Q=m_1=0$, which 
satisfies conventional associativity.

\newsec{Topological D-branes and the derived category}

As we discussed in the introduction, a complete category of
``topological D-branes'' must contain both branes and antibranes.
This might be naturally accomplished by introducing a $\BZ_2$ grading
on the boundary conditions.  However, the discussion of the previous
section shows that this is contained in a $\BZ$ grading of the
boundary conditions, and that for $\hat c>1$ it is better to keep
this, because it is related to the degree of morphisms.  (One might
instead keep a $\BZ_{2\hat c}$ grading, but there is no real advantage
to doing this.)

Let us therefore introduce the notation $E_n$ for a boundary condition
located at $\grade=n$, and the notation $E[n]$ for the shift of a boundary
condition $(E[n])_m = E_{n+m}$ (the ``translation functor'').
At the end of the discussion we will
regard the simultaneous translation of all objects by $E\rightarrow
E[1]$ combined with reversing all the K theory classes as a gauge
symmetry, but until we get there, we will regard these as distinct
objects.

As usual we can consider the direct sum of a set of boundary conditions
to be a new boundary condition, distinguished by ``Chan-Paton factors.''
Let us generalize the preceding notation: if $E$ is such a direct sum,
let $E_n$ be the component located at $\grade=n$.

A map between two such direct sums $E$ and $F$ is a direct sum of
components of definite $U(1)$ charge,
\eqn\mapcoll{
\Ext^p(E,F) = \oplus_{n,k} \Ext^{p-k}(E_n,F_{n+k}) .
}
This formula includes a boundary contribution to $U(1)$ charge,
by adding the ``distance'' between the images to the fermion number.

Since we now have $U(1)$ charge living on the boundary,
we can make a further generalization to put a boundary component in the
BRST operator, consistent with it having charge $1$.  Thus we write
$$
Q_{(E,F)} = Q^{(0)}_{(E,F)} + d_E - d_F
$$
with
$$
d_E \in \oplus_{n,k} \Ext^{1-k}(E_n,E_{n+k})
$$
and similarly for $d_F$.
The operator $Q^{(0)}_{(E,F)}$ is the ``original'' BRST operator
(not acting on the Chan-Paton factors) and if we take all of our
states to live in its cohomology, can be taken as zero (this could
conceivably be generalized).
In this case, the condition $Q^2=0$ will be satisfied if
$$\eqalign{
\{d_E,d_F\} = 0 \cr
d_E^2 =0 \cr
d_F^2 = 0.
}$$
The first equation is conventional, while the other two tell us that
$(E,d_E)$ and $(F,d_F)$ are ``complexes'' as defined in
homological algebra.  The $Q$-cohomology then consists of
maps $\phi$ in \mapcoll\ such that $d_F \phi = \phi d_E$
(chain maps) modulo the image of $d_F \epsilon - \epsilon d_E$
(homotopic equivalence).\footnote*{
Actually, complexes are usually defined in terms of $d$ containing only
degree zero maps
$\Hom(E_n,E_{n+1})$, and this is all we will need below.  The apparent
generalization will go away in the final result.}

\subsec{Adding brane-antibrane pairs}

The main thing we gain from all of this formalism is the ability to
describe branes and antibranes on an equal footing.  This requires
us to be able to identify an object $E$ with an object
$F$ related by adding cancelling brane-antibrane pairs.  

Now we want a finer equivalence relation than K theory: a pair $A\bB$
will only be considered to cancel if $A$ and $B$ are isomorphic as
holomorphic objects, in other words if there is an identity map $1$ in
$\Hom(A,B)$.\footnote{$\dagger$}{
One can make a similar refinement by
making the equivalence $[A]+[C]\cong [B]$ only when there exists an
exact sequence $0\mapr A\mapr B\mapr C\mapr 0$.  This leads to
analytic K theory, which will distinguish
different points in $M$, for example.  However, this does not give
the morphism or grading structure of the derived category, which will also
be important physically.}
We furthermore require that adding the pair does not
change the $Q$ cohomology: for every object $G$, we have $\Hom(E,G)
\cong \Hom(F,G)$.

This can be accomplished by adding a pair $B_n\oplus B_{n+1}$ with
grades differing by $1$, and taking the new BRST operator $Q_{F,G}$ to be the
direct sum of the old BRST operator $Q_{E,G}$ with the identity map 
$\bf 1$
on morphisms in $\Hom(B,G)$ involving the brane-antibrane pair:
$$
Q_{F,G} = Q_{E,G} \oplus {\bf 1} .
$$
This identity map $\bf 1$ is
the composition $\Hom(B,B) \times \Hom(B,G) \rightarrow \Hom(B,G)$.
If we had considered the $(G,F)$ open string sector, it would act as
$\Hom(G,B) \times \Hom(B,B) \rightarrow \Hom(G,B)$.
This term in $Q$ of course pairs each $\Hom(B_n,G)$ with a $\Hom(B_{n+1},G)$
and removes them from the topological Hilbert space.

By adding brane-antibrane pairs in various degrees to our original
objects, we can get a large class of chain complexes.  One can
understand the main properties of this construction by thinking of the
$E_n$ as vector spaces of Chan-Paton factors and the summands of $d$
as matrices.  Although at first it looks like we will only get very
special $d$ (made up from identity matrices), of course change of
basis (complex gauge transformation) and other operations will produce
more general complexes.

We now define our category of topological D-branes as the result
of identifying any pair of complexes which are related by the following
types of morphisms: the morphism which adds brane-antibrane
pairs discussed above, complex gauge transformations, morphisms
homotopic to the identity ({\it i.e.} those of the form $1+Q\epsilon$
which are equivalent to the identity in $Q$-cohomology),
and compositions of any of these.
Furthermore, we identify two objects $E$ and $F$ if for each there 
is a morphism in this class mapping it to a third complex $R$,
their ``common refinement.''  
The additional identifications we postulated are unavoidable
if we do not wish to distinguish objects related by an isomorphism.

There is one subtlety in correctly making these identifications
which we now explain.\footnote*{ 
This was stated incorrectly in the original version,
and I thank R. Thomas for pointing this out to me.}
In identifying two configurations which differ by adding
brane-antibrane pairs, it does not suffice to only identify
configurations related by the direct sum we just discussed;
a more general identification is required.

Physically, the point is that we want to identify $E$ with $E + X +
\bar X$ obtained by adding a cancelling brane-antibrane pair, even if
$E$ and $X$ (or $E$ and $\bar X$) themselves were combined in a bound
state, say $F$.
To explain the mathematical point, we quote the incorrect argument
from the original version, to see where it was wrong.
The mathematical description of forming
$F$ as a bound state of the branes $E$ and $X$ is a non-split
exact sequence:
$$
0 \mapr E \mapr^f F \mapr^r X \mapr 0
$$
By definition, the maps satisfy $r\cdot f=0$.  A split exact
sequence is one for which
$F\cong E\oplus X$, in which case $F$ is not
physically a bound state.  A bound state can be obtained from this by
turning on an off-diagonal bosonic mode, breaking the gauge symmetry
back to $U(1)$.  Such a mode will again be associated to a partner
fermionic open string between $E$ and $X$, and (by general mathematical
formalism we will come back to) to an element of $\Ext^1(X,E)$.

Now the point is that physics (crossing symmetry) requires that one
can also make a partial annihilation $F + \bar X \mapr E$, and
the same exact sequence must also describe this.  This will be true
if our identifications equate $E$ with the complex $F \mapr^r X$.
The physical identity of these configurations
can also be checked in examples, say in large volume.

One can try to do this using a homotopy equivalence $1+Q\epsilon$, and
this is possible if a partial inverse $\epsilon$ to $r$ can be found,
satisfying $r\epsilon = 1|X$.  However this is precisely the case in
which $F = E \oplus X$ is not really a bound state; the open string
mode is not turned on.

These considerations lead us to
identify all complexes which are related by general
quasi-isomorphisms.
A quasi-isomorphism is an element of $\Hom(E^\cpxdot,F^\cpxdot)$ (the
$Q$-cohomology in our previous language) which is an isomorphism when
restricted to cohomology.  This includes all the morphisms we
discussed, but only transformations which add pairs or sequences of
objects which cancel out of the cohomology, both of individual objects
and of morphisms between objects.

A quasi-isomorphic pair is a pair of
complexes each related by quasi-isomorphism to a common refinement;
all such pairs are identified.  
The result of identifying all quasi-isomporphic pairs
is the derived category $D(\CA)$
of the category $\CA$ we started with -- in our present discussion
$\Coh X$, the coherent sheaves on the CY $X$.
As discussed in \gm, this is usually
done by localization, i.e. one allows as morphisms formal inverses of
all the quasi-isomorphisms, and then shows that one can write any
product of morphisms in terms of a single morphism by commuting these
inverses through other morphisms (or ``combining denominators'').  
The result is an associative category, though not abelian (kernels and
cokernels need not make sense).

All this may be made clearer by an explicit
example, which we will give in section 5.

\subsec{Resolutions and $\Ext$}

Resolutions are a key tool in homological algebra, and will turn out
to give us a mathematical counterpart of the ``images'' we introduced
in the previous section.

A free resolution of a complex $E$ is a quasi-isomorphic
complex $R$ in which the terms are free.
In the context of vector bundles on complex projective space $\CP^n$,
this means they are direct sums of line bundles.
The point of this is that it means that there
are no relations hidden in the definition of the terms; all the relations,
and thus the homology, are explicit in the maps $d_R$.

Free objects or modules are too restrictive for most purposes and
one usually discusses the more general concepts of injective and projective
resolution.  A projective object is a direct summand of a free object,
while an injective object can be defined in terms of this by dualizing
and reversing arrows.  Considerations we mention shortly lead to a more
useful criterion: a projective object $P$ has 
$\Ext^i(P,X)=0\ \forall X$ and $i>0$, while an injective object $I$ has
$\Ext^i(X,I)=0\ \forall X$ and $i>0$.

An injective resolution of $F$ is an exact sequence
\eqn\injres{
0\rightarrow F\mapr^{f} I^0\mapr^{r_0} I^1\mapr^{r_1} I^2\mapr \ldots
}
where the $I^n$ are injective, and 
one can similarly define projective resolutions 
$$\cdots\mapr P^1 \mapr P^0\mapr F\mapr 0$$
where the $P^n$ are projective.  

It can be shown that every coherent sheaf on a space $M$ has an
resolution whose terms are locally free sheaves on $M$, 
i.e. sheaves which in any local region are equivalent to bundles,
and thus the use of
resolutions allows reducing general computations to those involving bundles.
In particular, they can be used to
define sheaf cohomology or better its generalization, the $\Ext$ groups.
We can define $\Ext^k(E,F)$ as the cohomology $H^k$ of of the complex
\eqn\homcomplex{
0\mapr \Hom(E,I^0)\mapr \Hom(E,I^1)\mapr \Hom(E,I^2) \mapr \ldots
}
or equivalently as homotopy classes of chain maps from $E$ into the
resolution.  Using the equivalence of $F$ with its resolution
in the derived category, this also tells us that
$$
\Ext^p(E,F) \cong \Hom(E,F[p]) .
$$
In other words, an $\Ext^p$ is a degree zero map into the $p$'th term
of the resolution.  This allows us to think of the successive terms in
the resolution as providing a concrete picture for the ``images'' we
introduced in the previous section.

We define the {\it length} of a complex to be the number of non-zero
terms $E_n$ minus one.  For each object, there is a minimal length of the
complex required for its free resolution.  The maximal such length for
a given category is the homological dimension $hd$ of the category;
for coherent sheaves on a complex manifold this will generally be the
dimension of the manifold (it can be less).  This is easy to see for
bundles if we assume the relation 
$\Ext^k(E,F)\cong H^k(M,E^*\otimes F)$ and consider \homcomplex.

The relation between the length of the complex and the highest $\Ext$
group which can appear has another consequence, namely that one cannot
get all sheaves on a $d$-fold by using resolutions of length less than
$d$.  In particular, the monad construction (which is closely related
to this) with a complex of length two and with line bundles as
constituents cannot describe all bundles on a generic Calabi-Yau
three-fold.  It would be interesting and probably quite useful if
linear sigma models could be generalized to use longer complexes.

\subsec{Properties of the derived category of coherent sheaves}

As we mentioned in the introduction, the derived category of coherent
sheaves contains far more information than the K theory.  An
illustration of this is the reconstruction theorem of Bondal and Orlov
\bondalorlov: in certain cases (with ample or anti-ample canonical bundle;
this excludes Calabi-Yau), the variety $X$ is determined by $D(\Coh
X)$.  The strategy is to first identify the set of objects
corresponding to points; the morphism information can then be used to
put a topology on this set and show that it is indeed the expected
variety.

Although this result is not literally true for a Calabi-Yau, the
reason that it fails is quite interesting and relevant for us.
It is that there is not a unique definition of which objects are the
points.  Indeed, there are ``autoequivalences'' of the derived category --
transformations which permute the objects but preserve the structure of
the morphisms -- which turn the points into other sheaves (or complexes).

These autoequivalences have been much studied in recent mathematical
work and all of them can be obtained as Fourier-Mukai transformations (FMT's).
A FMT from sheaves on a space $X$ to sheaves
on a space $Y$ (possibly the same as $X$)
is defined by specifying a sheaf $\CF$ on $Y\times X$ satisfying certain
properties; most notably, the restrictions to two points on $X$
$\CF_{x_1}$ and $\CF_{x_2}$ must satisfy 
$$\eqalign{
\Hom(\CF_{x_1},\CF_{x_2}) &\cong \BC \qquad {\rm if}\ x_1=x_2 ; \cr
&= 0 \qquad {\rm if}\ x_1\ne x_2.
}$$
The transform of a sheaf $E$ on $X$ is then
\eqn\fmt{
\hat E = R\pi_{Y*} (\CF \otimes^L \pi_X^* E).
}
The idea expressed by this formula is simple and well explained in \thomas:
one pulls back $E$ to the product space, tensors
with $\CF$, and then ``pushes forward''
in the sense that the resulting sheaf can have as local sections any of 
the local sections of the product sheaf 
(this construction is referred to as the ``direct image''),
but with the dependence on $X$ is suppressed -- this is the reason for the
name ``Fourier'' as taking all such sections is like integrating over $X$.

The detailed implementation of this idea requires resolving the
sheaves which appear in intermediate steps (this is the meaning of the
``$L$'' and ``$R$'' symbols), and is greatly simplified by working
with the derived category.

A particularly simple set of FMT's are the ``twist functors''
discussed in detail in \seidelthomas.  For every sheaf $E$ on $X$
there is a twist functor $\CT_E$, which has all the right properties
to correspond physically to a monodromy associated to a loop in
K\"ahler moduli space around a point at which $Z(E)=0$, such as a
conifold point.  Assuming this is so, all of these monodromies
preserve the derived category of topological branes and this is fairly
strong evidence that any physical construction of a model associated
to a CY with a specific complex structure, will produce the same
derived category of objects.

We will discuss this in more detail elsewhere; here we will
motivate the theorem of Beilinson used in \noncompact\ using these ideas.
This states that the derived category of sheaves on $\P^n$ is equivalent
to the derived category of representations of the quiver-complex
$QC(n+1,n+1)$ (to be defined in section 5), with $Y=0$.

This comes from a one-to-one correspondence between sheaves, implemented
by an FMT in which $\CF$ is a sheaf on $\P^n \times \P^n$
which is just the ``delta function'' (or structure sheaf) supported on
the diagonal.  Such an $\CF$ will clearly produce the identity
transformation on $E$.

The non-trivial content of this construction comes when we look at the
resolution of this delta function sheaf.  This is the
``Koszul resolution'' which is a complex with successive terms
$$
\Lambda^n\Omega_Y(n) \times \CO_X(-n) \mapr \ldots \mapr
\Omega_Y(1) \times \CO_X(-1) \mapr \CO_Y \times \CO_X \mapr \CF.
$$
The tensor product appearing in \fmt\ is 
$$
\CF \otimes E = 
\Lambda^n\Omega(n) \otimes E(-n) \rightarrow \ldots \rightarrow
\Omega(1) \otimes E(-1) \rightarrow \CO \otimes E
$$
and continuing in this vein, one finds that the FMT (which is
equivalent to the original object in the derived category) is a
complex of the sheaves $\Lambda^p\Omega_Y(p)$ tensored with the
cohomologies of the original sheaf $H^m(E(-p))$.  If these
cohomologies are non-zero in a single degree $m$, they can be thought
of as defining an object in $QC(n+1,n+1)$, providing the
correspondence used in \noncompact, while if they are non-zero in more
than one degree, one gets an object in $D(QC(n+1,n+1))$.

\newsec{K\"ahler moduli and flow of grading}

Having understood the category of boundary conditions in the topological
string theory (or ``topological D-branes''), we now can assert with
confidence that every physical D-brane corresponds to a unique topological
D-brane.  We now want to understand why not every topological D-brane
corresponds to a physical D-brane.

In particular, let us explore what happens if we start with a BPS
configuration of two branes and then vary the K\"ahler moduli.  In
general, the grades of the two branes, $\grade_1$ and $\grade_2$,
will no longer be equal.

Although open string sectors with $\grade_1\ne\grade_2$ bear some
resemblance to twisted $\CN=2$ sectors, they are not the same.
Because a specific $\CN=1$ world-sheet supersymmetry is gauged in the
open superstring definition, this supersymmetry must still have
conventional NS and R boundary conditions on the $\CN=1$
supercurrents.  Furthermore, it will still admit a GSO projection; two
sectors related by a continuous deformation will share the same GSO
projection.  This is possible because both bosons and fermions in
these sectors will have their moding changed; the fermions in a way
determined by \specbc\ and the bosons in a corresponding way to make
NS and R supercurrents possible.  A solvable example in which this can
be seen is the theory of two $3$-branes oriented at angles \bdl.

As discussed earlier, if we restrict ourselves to K\"ahler variations,
the massless fermion sector will remain unchanged.
Now even though these combinations of boundary conditions have broken
supersymmetry, we can still identify bosonic partners of the
massless fermions, as the NS sectors accessible by varying the $U(1)$
charge (spectral flow).  Another way to say this is that since the
individual boundary CFT's correspond to BPS branes, they each have
spectral flow operators, and we can use either of these to define the
action of space-time supersymmetry.  Depending on which one we use,
we will get different results, but these will only differ by a phase.
Thus we can still identify a unique NS state as the superpartner.

Let us discuss the various physical states in the CFT we obtain 
by this construction, and their space-time interpretation.
We start with NS states in a sector with $\Delta\theta=0$ and the usual
$\CN=1$ supersymmetry.  Let $Q$ be the $U(1)$ charge in the $\CN=2$
algebra; then $Q=q=0$ states are gauge bosons (the standard GSO
projection will keep only states $\psi^\mu_{-1/2}\ket{}$ in the
space-time CFT) while $Q=q=\pm 1$ states are bosons in chiral
multiplets ($q=-1$ will be the complex conjugates of the $q=+1$
states).

Varying the relative grading will shift
the $U(1)$ charge of all of the states in this list, and preserve the
GSO projection.  Our conventions are such that the charge shift is equal
to the shift in grading, $\Delta q = \Delta(\grade_1-\grade_2)$.

Let us write the Ramond vertex operator as a product of internal,
bosonized $U(1)$ and space-time factors
$$
O_\theta\ e^{i (q-3/2) \phi}\ V_{3+1} .
$$
As discussed above, varying $\grade$ by only varying
the K\"ahler class keeps the Ramond state massless, and
this means that the dimension of
$O_\theta$ is determined by $h(O_\theta)+(q-3/2)^2/6=3/8$.
We can then derive the dimension of the operators related to it by
the usual $\CN=2$ spectral flow and thus the mass squared of the
partner NS states with $U(1)$ charge $q$.  These will be
\eqn\bosonmasses{\eqalign{
m^2 &= {3\over 8} + {q^2\over 6} - {(q-3/2)^2\over 6} - \half \cr
&= \half (q-1)
}}
for chiral multiplets.
A vector multiplet has $V_{3+1}=\psi^\mu$,
so $m^2=q/2$.

The bosons in chiral multiplets come from $q=1$ states, and we see
from this that $\Delta\grade$ enters into their mass squared precisely as
a Fayet-Iliopoulos term  would.  This is the CFT argument for the
earlier description of BPS decay by D-term supersymmetry breaking
\refs{\sharpe,\kamg,\pistable}\ and indeed these masses can be often
be modelled by assigning the FI terms associated to the $U(N)$ space-time
gauge groups of the two groups of branes the values $\zeta_1=\grade_1$
and $\zeta_2=\grade_2$.  Starting from an $\Ext^1$ (a massless chiral
multiplet), this assignment reproduces \bosonmasses.  This explicit
$\CN=1$ field theory picture has its limitations however as seen in
\dd\ and it is for this reason that, although the formalism and results 
are very similar to those in $\CN=1$ supersymmetric field theory, we
have not based our discussion on this similarity but instead on CFT.

We should also mention at this point that in our setup, the precise
definition of a theory with tachyonic strings between a pair of branes
is to take branes which do not fill Minkowski space (e.g. D0-branes)
and separate them in these dimensions.  This gives the tachyons large
masses and makes the configuration with zero tachyon vev stable, while
still allowing us to argue that such a configuration of coincident
branes would be unstable.

Starting from $q=1$ and making such a flow, $m^2$ can decrease until
we reach $q=0$.  At this point, the operator $O_\theta$ has dimension
zero.  Because the world-sheet theory is a (correlated) product of a
unitary theory with the bosonized $U(1)$, zero is the minimal possible
dimension for $O_\theta$.  Thus, if we decrease $q$ further, our
assumptions must break down.  One also knows from $\CN=2$ representation
theory that chiral operators must have $0\le q\le \hat c$ 
(for the same reason).

Now the boundary states we are discussing were unstable as soon as we
had $q<1$ and a tachyonic open string; we interpret these arguments as
showing that nevertheless we can treat them as sensible boundary
states and use them to define a category of boundary states as
discussed earlier, but only as long as $q\ge 0$.  If $q<0$, the massless
Ramond state no longer makes sense in unitary CFT, and must disappear
from the theory.  Note that it is not becoming massive (in general there
is no state for it to pair with) but literally disappearing, which can only
happen if one or both of the boundary states also disappear from the theory.

This is the essential subtlety in the relation between topologically
twisted open string theory and the physical open string theory.  From
the point of view of topological theory, there is nothing wrong with
these boundary states (since we varied only K\"ahler moduli, nothing
changed).  One can even keep track of the $U(1)$ charges in this
situation: the same additive conservation laws will hold even though
some $U(1)$ charges may be negative.  However, any given physical
CFT can only realize a subset of these boundary states, chosen
so that morphisms of negative degree do not appear.

From our arguments, the choice of which boundary states are realized
would be expected to depend on K\"ahler moduli.  Starting from an
$\Ext^1$ between a pair of simultaneously BPS branes, one can reach
$q=0$ if the relative phase difference reaches $\pi$, i.e. if the BPS
central charges anti-align.  In this sense, the phenomenon
under discussion has to do with branes turning into anti-branes.

\subsec{Special Lagrangian picture}

Many of the statements we just made can also be seen from the geometry
of the A brane picture.\footnote*{ I am indebted to Richard Thomas and
Paul Seidel for explaining this to me.}  
The analog of a coherent sheaf in this picture is an isotopy class of
Lagrangian manifolds, while a physical brane is a particular
Lagrangian in this class satisfying the ``special'' condition for the
chosen complex structure and holomorphic $3$-form.  The morphisms are
given by Floer cohomology.

The standard convention for the grading on Floer cohomology agrees with
our notations $\Hom$ and $\Ext^1$ for vector and chiral multiplets
respectively.  To determine this grading, one looks at
the three relative angles $\Delta\theta_i$.  There are basically two
cases: where these all have the same sign (producing a $\Hom$),
or where one sign differs from the other two (producing an $\Ext$).
This rule agrees with the physical GSO projection in the branes
at angles model of \bdl.

In these terms, our grading is $\Hom(A,B[n])$ or $\Ext^1(A,B[n])$
with $n=\sum_i\Delta\theta_i$.
As one varies the complex structure, the $\Delta\theta_i$ will
vary, leading to the flow we just described.  In this picture, the reason
the gradings cannot flow below zero (or above $\hat c$) is that given a
pair of Lagrangians which intersect transversely, they will continue
to intersect transversely under any change of complex structure.

The conclusion is the same, that gradings will flow, but cannot
flow below zero for physical objects.

\subsec{Some convenient notation}

As we motivated earlier (in terms of resolutions and the derived
category), we will denote a morphism 
with $U(1)$ charge $n$, i.e. a fermionic string
between the brane $A$ and the $n$'th image of the brane $B$,
as an element of $\Hom(A,B[n])$.
This group depends only on the difference between the grades:
$\Hom(A,B[n]) \cong \Hom(A[m],B[n+m])$.

In addition to its other mathematical advantages,
this notation makes it simple to 
express the variation of the $U(1)$ charge of the morphism
as we vary the K\"ahler moduli.
It simply comes from varying the grades of the two branes:
$\Hom(A,B[n])$ will flow to 
$\Hom(A[\Delta\grade(A)],B[n + \Delta\grade(B)])$.

We now introduce a notation which allows us to keep track both
of our new grading, and the ``original'' grading (that present in
Floer cohomology), since both are needed in string theory.
We distinguish the massless fermions leading to chiral and vector
multiplets, by writing a degree $n$ morphism as
$\Hom(A,B[n])$ for a gaugino and $\Ext^1(A,B[n-1])$ (or just $\Ext$)
for the fermion in a chiral multiplet.
(One motivation for this notation is the observation
that the partner boson has $m^2$ proportional to the quantity in
brackets in both cases.)

The two groups $\Hom$ and $\Ext$ as we defined them
are related by double spectral flow, which reverses the
GSO projection.  This tells us that 
$\Hom(A,B[n]) \cong \Ext(A,B[n+\hat c-1])$
and thus $\Ext$ is not really independent.

We define the antibrane $\bar A$ 
to a brane $A$ by reversing the GSO projection, so
$\bar A \cong A[\hat c]$.  Any simple brane $A$ (with gauge group
$U(1)$) will have $\Hom(A,A)\cong\BC$, and we can infer the existence of
the brane-antibrane tachyon $\Ext(A,\bar A[-1])$ from this.
More generally one has $\Hom(A,B[n]) \cong \Ext(A,\bar B[n-1])$.

In the large volume limit, we can use Serre duality to infer the
existence of additional morphisms.
In the CY theories of prime interest for us, this takes the form
$\Hom(A,B[n]) \cong \Hom(B,A[\hat c-n])^*$.
This is a dual relation, not an identification,
but can of course be used to relate the dimensions of
these spaces.  Its most fundamental role in this discussion is that
it is the inner product of the topologically twisted open string theory
(the shift $\hat c$ is the anomalous $U(1)$ charge of the disk).

In our application, this duality also reverses the GSO projection.
This is because $\hat c$ is odd, so the Serre dual morphism will have opposite
parity $U(1)$ charge, and we infer the GSO projection directly from this.
This leaves us with the rule
\eqn\dualrule{
\dim \Hom(A,B[n]) = \dim \Ext^1(B,A[\hat c-n-1]) .
}

These rules determine the gradings of morphisms at arbitrary points in
K\"ahler moduli space given the gradings at one point.  It is natural
to start with the large volume limit as this point.  One should note
that gradings of morphisms between branes of different dimension
(i.e. sheaves with support of different dimension) in our conventions
do not agree with the usual grading of cohomology.  We can get them
either by carefully computing $U(1)$ charges, or masses of partner
bosons in CFT.  The latter generally can be obtained from standard
D-brane considerations \pol; in particular the lightest NS string
between a D$p$-brane and a D$q$-brane has $m^2=|p-q|/8-1/2$, from the
shift of the ground state energy due to ``DN bosons.''

This gives a morphism from a D$p$-brane to a D$q$-brane degree
$n+(p-q)/4$ (one must be careful about orientations in identifying
both objects as ``branes'').  Taking this rule into account along with
the large volume asymptotics for the periods, one finds that morphisms
between simultaneously BPS branes (those whose grades differ by
integers) will always have integer grading.

Another large volume subtlety we should mention is that one needs to
be careful to define the grade in a way which depends smoothly on the
Chern classes (the branch of the logarithm cannot jump).  This in
particular corrects an observation made in \noncompact\ -- by doing
this, one can get the large volume limit of $\Pi$-stability to
reproduce Gieseker stability in all cases (this comes from the dependence 
of subleading terms on the higher Chern classes).
This is important as otherwise one finds many incorrect predictions --
for example, ideal sheaves of points will destabilize the trivial
line bundle unless one has Gieseker stability.

\newsec{Examples of flow of grading}

Two classes of examples have been studied in some detail: orbifolds
$\BC^3/\BZ_K$, and Gepner models or Landau-Ginzburg orbifolds, which in 
a certain sense are hypersurfaces in $\BC^5/\BZ_K$ orbifolds.
We will not review here the discussion of
\refs{\bdlr,\dg,\noncompact,\dd}\ which leads to the following identifications
of boundary states and quiver theories but only cite the results.

For the most recent work on Gepner models, see
\refs{\govgep,\mayrgep,\tomgep}.

\subsec{Kronecker quiver theories}

Besides defining the quiver theories which will appear in our
examples, the next two subsections define the quiver $\Ext$ groups,
which will be compared with their large volume analogs, and
illustrates how methods of homological algebra can be helpful in
analyzing their moduli spaces of supersymmetric vacua.

Let us consider a theory with gauge group
$U(n_1)\times U(n_2)$, and $q$ chiral multiplets $X^a$
in the  bifundamental $(\bar n_1,n_2)$.
The $X^a$ can be regarded as matrices or maps $X:\BC^{n_1}\mapr\BC^{n_2}$.
Complex gauge equivalence classes of configurations will be referred
to as objects in the category $QC(2,q)$.

The simplest operation we can define for these objects is direct sum,
and this allows us to define K theory in the usual way in terms of
pairs of objects.  Any ``topological'' definition of K theory class
will not see the configuration (the $X$ and $Y$'s can be continuously
deformed to zero); thus the K theory class of an object is just the
set of integers $n_i$.

It is also clear that the K theory class is only the most basic
invariant of an object; in general the objects come in moduli spaces
of finite dimension.  For $p=2$, there is an obvious formula for the
dimension of this moduli space, obtained by counting matter fields
modulo gauge symmetries.
If the object is simple, i.e. breaks the
complexified gauge group to $GL(1)$, it is
\eqn\naivedim{
d = 1 + q n_1 n_2 - n_1^2 - n_2^2 .
}

Let us write a slightly more general formula from which this
formula can be deduced but which is valid in general.
Given two objects $E$ and $F$,
we define an element of
$\Hom(E,F)$ to be a pair of maps 
$\phi_i:\BC^{n^E_i}\mapr\BC^{n^F_i}$, $i=1,2$,
satisfying
$\phi_2 X_E^i = X_F^i \phi_1$.
An element of $\Ext(E,F)$ is a map 
$\phi^{(1)}:\BC^{n^E_1}\mapr\BC^{n^F_2}$ which cannot be written
as $\phi^{(1)}=\phi_2 X_E^i - X_F^i \phi_1$ for some $\phi_1,\phi_2$.

One can easily check that $\Hom(E,F)$ are off diagonal gauge transformations
which are unbroken in the configuration $E\oplus F$, while $\Ext(E,F)$
are off diagonal chiral multiplets which cannot be gauged away in this
configuration.  Thus the dimension of moduli space we discussed earlier
is the special case $d=\dim\Ext(E,E)$.

One can then write
$$
\chi(E,F) \equiv \dim\Hom(E,F) - \dim\Ext(E,F) =
n_1^E n_1^F + n_2^E n_2^F - q n_1^E n_2^F ,
$$
a formula which can be proven
by showing that $\chi(M,N)$ is ``topological'' (invariant
under deformations of the $X$'s), so that one can compute it for $X=0$.
This generalizes in an obvious way to any quiver theory with no
superpotential.

It is easy to work out the homological algebra of section 3
for these theories.
A free object is one with a basis, i.e. the elements
$X^i e_j$ with $e_j$ in $\BC^{n_1}$ must be linearly independent.
This requires $n_2 \ge q n_1$ and puts a condition on the $X^i$.
The simplest example is $(1\ q)$, say with $X^i_{1,j}=\delta^i_j$.

A projective object is a direct summand in a free object.
In the example, the only
indecomposable projective objects are
the two objects $(0\ 1)$ and $(1\ q)$ as above.

The projective resolution of $(1\ 0)$ is then
$$
0 \mapr (0\ q) \mapr (1\ q) \mapr (1\ 0) \mapr 0.
$$
This does not come from the direct sum
adding $(0\ q)\mapr (0\ q)$ as this would
have given $(1\ q)$ with $X^i=0$, which is not projective.
Nevertheless one must identify this object with its resolution to get
sensible mathematics and physics.

\subsec{Quiver-complex theories}

Let us now consider a theory with $p$ gauge groups,
$U(n_1)\times U(n_2)\times\ldots\times U(n_{p})$, and a global $U(q)$
symmetry.  The matter spectrum is bifundamentals $X_i^a$
in the $(\bar n_i,n_{i+1})$ for each $1\le i\le p-1$ and in the fundamental
of $U(q)$, and bifundamentals $Y_i^{[ab]}$ in the $(\bar n_{i+2},n_i)$
for $1\le i\le p-2$ in the antisymmetric $[q-2]$ representation of $U(q)$.
For $p>2$ there is a superpotential,
\eqn\qcsuper{
W = \sum_{i=1}^{p-2} \tr X^a_{i+1} X^b_i Y^{[ab]}_i.
}
We will consider the 
category $QC(p,q)$ whose objects
are (classical) solutions of the F-flatness constraints for
all of these theories.

It turns out that there is no such universal formula which gives the
dimension of moduli space just in terms of the K theory class for
$p>2$.  (There is a formula which works in almost all cases for $p=3$
and $q\le 3$, but not for higher $p$ or $q$.)  Well-known arguments
that such a formula is not to be expected in general $\CN=1$ theories
are the possibility of lifting arbitrary pairs of chiral multiplets by
changing the configuration, and the existence of cases in which the
moduli space has several branches of different dimension.

One can however go further than these general statements by better
understanding the problem.  Given a superpotential, one might try to
predict the dimension of the moduli space by subtracting the number of
relations following from $W'=0$ from the prediction \naivedim.  Since
there is one relation for each matter field, this always leads to a
negative dimension and would predict that no solutions exist.
In fact solutions can exist, but only when the relations are redundant;
these redundancies will depend on the specific configuration.

The redundancies between relations in the quiver complex theory
can be understood in homological terms.  We start
by considering only the configurations with $Y=0$; the relations are then
$$
X_{i+1}^a X_i^b = X_{i+1}^b X_i^a
$$
which can be expressed (for $q\ge p$) as $d^2=0$ on a complex of
vector spaces.  This complex can be defined in terms of an exterior
algebra $\Lambda V$ where $V\cong \BC^q$ has a basis $e_a$ which we think
of as anticommuting objects: $e_a e_b + e_b e_a = 0$.
The operator $d$ then acts on an element of
$E_i\equiv \BC^{n_i}\otimes \Lambda^{i-1} V$ to produce
an element of $E_{i+1}$ as multiplication by $X^a e_a$.

One then defines the ``morphism complex'' as follows.  A ``chain map''
$\phi^{(n)}$ of degree $n$ from $E$ to $F$ is a set of linear maps 
$\phi^{(n)}_i$
from $E_i$ to $F_{i+n}$ which can be written in terms of the $e_a$;
i.e. 
$$
\phi^{(n)}_i = \phi^{(n)}_{i,{a_1}\ldots{a_n}} e_{a_1} \ldots e_{a_n}.
$$
One can then regard $d_F - d_E$ as an operator acting on these chain maps;
it squares to zero, so we can define
a morphism of degree $n$ as an element of its cohomology.
The space of such morphisms will be denoted $\Ext^n(E,F)$ (or
$\Hom(E,F)$ for $n=0$).\footnote*{
One can show that this is equivalent to the more general
definition involving resolutions made in section 3.}

All of this is very parallel to the discussion we made in section 3
and the reader may be wondering why we repeated it.  The main reason is
that while formally it is very parallel, physically its interpretation
is rather different: in particular, we do not make the identifications
we made in section 3, but instead interpret configurations just as
we would in supersymmetric gauge theory (identifying only configurations
related by complex gauge transformation).  The result is (at least
mathematically) an abelian category to which one can then apply the
construction of section 3, considering complexes of these complexes
and forming the corresponding derived category.  
This is the type of derived category which according to Beilinson's theorem
will be equivalent to derived categories of coherent sheaves.

Returning to concrete considerations, these definitions lead immediately to
a formula for the relative Euler character:
\eqn\euler{
\chi(M,N) \equiv \sum_{k=0}^{p} (-1)^k \dim \Ext^k(M,N) = 
\sum_{1\le i\le j\le p} (-)^{i-j} \left({q\atop i-j}\right) m_i n_j.
}
This is proven in the same way as before.

Before going on to the generalization with $Y\ne 0$, let us discuss the
physical meaning of this construction.  It is not hard to see that elements
of $\Hom(E,F)$ correspond to (complex) gauge symmetries which appear when
we combine the theories $E$ and $F$ (the cohomology condition says that the
matter configuration is preserved by this gauge transformation).
Similarly, elements of $\Ext^1(E,F)$ correspond to massless matter multiplets
(or linearized moduli): elements of cohomology correspond to deformations
which are not lifted by the superpotential and are not pure gauge.

The higher $\Ext^n$'s do not admit such an obvious interpretation but clearly
have to do with relations between the superpotential relations; when these
exist, one cannot derive the dimension of moduli space, which
is $\dim \Ext^1(E,E)$, just from the formula \euler: one needs more information
(such as the dimensions of the higher $\Ext$ groups).

In the case at hand, there is a second physical interpretation of
$\Ext^2(E,F)$: it counts massless deformations of the $Y$
multiplets.  This can be seen by considering the ``physical'' inner product
on $\Ext^n(E,F) \times \Ext^n(F,E)^*$ defined as
$$
(\phi,\psi) = \sum_{i=1}^{p-n} \tr \phi^+_{i+n} \psi_i 
$$
and the adjoint to $d$ defined by
$$
(\phi^{(n)},d^* \psi^{(n+1)}) = (d\phi^{(n)}, \psi^{(n+1)}).
$$
Now the F-flatness conditions on $Y$ are linear, $X^a Y_{[ab]} =
Y_{[ab]} X^b = 0$, and in fact are just $d^* Y = 0$.  The relation to
$\Ext^2$ then comes from the usual Hodge-style arguments that we can
find a unique representative of each cohomology class satisfying $d^*
Y = 0$ and that all solutions of these equations arise in this way.

This result is very suggestive of the relation 
$\Ext^2(E,F)\cong \Ext^1(F,E)^*$ valid on a Calabi-Yau threefold, and
indeed one expects such a relation to hold for any quiver theory which
arises in this context.  Morally speaking, the superpotential
\qcsuper\ is a finite dimensional
analog of the holomorphic Chern-Simons functional.

Indeed, as explained in \refs{\noncompact,\dd}, the case $p=q=3$ is
the $\BC^3/\BZ_3$ quiver theory, while $p=q=5$
describes sheaves on the quintic CY, and the other CY theories which
arose in \dd\ can be treated similarly.  Unlike the analogous formula
for sheaves on a CY$_3$, $\chi(E,E)$ can be non-zero; this is because
one has separately described deformations which made sense on the
ambient projective space (the $X$'s) and those which appear on
restriction (the $Y$'s) \ddtwo.  However, since $\dim \Ext^2$
enters into \euler\ with the wrong sign, one still needs more
information than $\chi(E,E)$ to compute the dimension of the moduli
space.  
(In special cases, the situation can be simpler; for example in 
the $\BC^3/\BZ_3$ quiver one can show that 
$\Ext^2(E,E)=0$ except for the D$0$-brane.)

This is made particularly clear by considering examples in which
the moduli space has branches of different dimension; these branches will
differ in $\dim\Ext^2$.  A simple example of this type is the quiver
with $n_1=1$, $n_2=2$ and $n_3=1$ in $QC(3,5)$ which appears as a rational
boundary state $\ket{11000}$ in the quintic Gepner model \bdlr; 
this moduli space has branches of dimension
5, 7 and 11, distinguished by $\dim\Ext^2 = 0$, 1 and 3.
The last of these must describe the rational boundary state.

These considerations define the morphisms in the quiver categories
we are about to discuss; their gradings are always CFT $U(1)$ charges,
which agree with the gradings we just defined for the $\BC^3$ orbifold
example (and in general if the fractional branes are simultaneously BPS),
but must instead be taken from CFT in the Gepner model example (since
the fractional branes are not simultaneously BPS).

We went into more detail than was required for this, to make the point
that these are concrete examples which capture the complexities of
categories of sheaves on Calabi-Yaus (and indeed are equivalent to
them once one goes to the derived category), but for which computing
the dimensions of homology groups (and thus of the local moduli space)
is just a problem of linear algebra.  Thus such problems, while not as
easy as the cases with more supersymmetry, are by no means
inaccessible.

\subsec{The $\BC^3/\BZ_3$ orbifold}

This leads to a quiver-complex theory
of type $QC(3,3)$.  At large volume the orbifold is resolved to
$\CO_{\P^2}(-3)$ and the three elementary or fractional branes can
be identified  \dg\ with the bundles (always on $\P^2$)
$B_1\equiv \CO(-1)$, 
$B_2\equiv \bar \Omega(1)$ (the antibrane to the twisted cotangent bundle)
and $B_3\equiv \CO$.

We first discuss the morphisms at the orbifold
point.  The quiver theory implicitly tells us what these are;
since all nine chiral multiplets are massless we see that
$\dim \Ext^1(B_i,B_j) = 3 \delta_{j\equiv i+1 (3)}$.

We next discuss the morphisms at large volume.
We can use standard methods to compute morphisms from sheaves to
sheaves, and then use the rules in section 4.2 to infer the morphisms
to ``anti-sheaves'' such as $B_2$.
The elementary morphisms correspond to
multiplication by the homogeneous coordinates: we have
$\dim \Hom(B_1, B_3) = 3$ and (using the Euler sequence)
$\dim \Hom(B_1,\bar B_2) = \dim \Hom(\bar B_2, B_3) = 3$.
Using the rule that $\Hom(A,B)\cong \Ext^1(A,\bar B[-1])$ we can
also say that
$\dim \Ext^1(B_1, B_2[-1]) = \dim \Ext^1(B_2, B_3[-1]) = 3$.
These ``brane-antibrane'' pairs naturally come with tachyonic open
strings.
Finally, the morphisms $\Hom(B_1,B_3)$ have Serre duals, which
according to our previous discussion are $\Ext^1(B_3,B_1[2])$'s.

\ifig\periodplot{The evolution of the central charges $Z(B_i)$ for
fractional branes in $\BC^3/\BZ_3$ from large volume, where they
asymptote to $Q_4(B+iV)^2$, to the orbifold point where they are
all equal.}{\epsfxsize2.0in\epsfbox{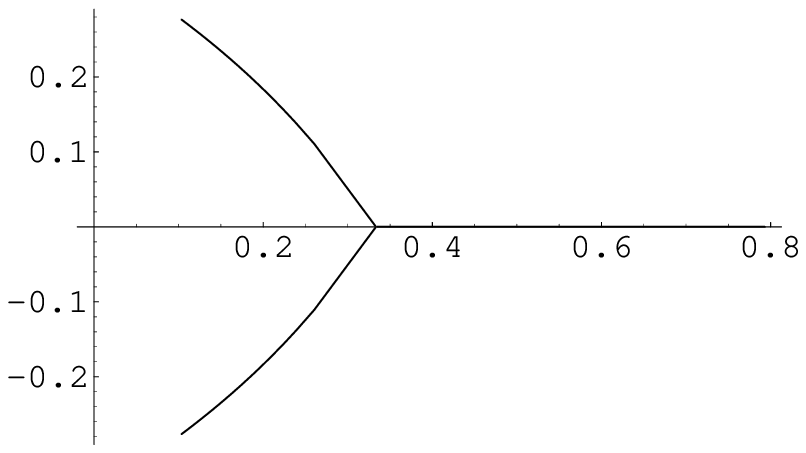}}

We can now check that the two limits are related under
the flow from large volume to orbifold point.
Referring to \periodplot, we see that 
$B_1 \rightarrow B_1[-1]$,
$B_2 \rightarrow B_2$, and
$B_3 \rightarrow B_3[1]$, and all of the morphisms we discussed work.
We also see that the superpotential satisfies topological
charge conservation, as the sum of the gradings ``around the triangle'' is
always 3.

One wants to check that all the morphisms agree, not just these
defining ones.  This should follow from the equivalence of derived
categories  established in \bkr, which is between representations of the
$\BC^3/\BZ^3$ quiver and sheaves with compact support
on the resolution $\CO_\P^2(-3)$.
The relation discussed in \noncompact, between the quiver with some
links set to zero and sheaves on $\P^2$, is a subset of this.

From a physical point of view, the difference between sheaves with compact
support on $\P^2$ and on $\CO_\P^2(-3)$ 
is that a D-brane wrapped on $\P^2$ in the second theory
has an extra world-volume field corresponding to the coordinate
transverse to $\P^2$, i.e. a chiral multiplet taking values in 
the line bundle $\CO_\P^2(-3)$, and the corresponding additional morphisms.
Thinking about the D-brane world-volume as
a topologically twisted theory \bsv\ shows that the fermion in this multiplet
lives in $\Ext^1(E,F \otimes \CO_{\P^2}(-3))$.  Another way of deducing these
morphisms is by using Serre duality twice, first on the total space and then
on $\P^2$, which on a $\Hom$ amounts to tensoring with this one-form.
Because of this, Serre duality on $\P^2$ also leads to relations on the
morphisms.

As an example (coming from \noncompact), we cite $\Hom(\CO,\CO(-3))$
which is non-zero at the orbifold point.  Flowing this back produces
$\Hom(\CO[-1],\CO(-3)[1])$ in large volume which indeed agrees with
$\dim H^2(\CO(-3),\P^2) = 1$.  The Serre dual to this is the
$\Ext^1(\CO_\P^2,\CO_\P^2(-3))$ we just discussed.

\subsec{Gepner models}

A product of five $\CN=2$ minimal models $A_{k_i}$ with $3=\sum_i
1-2/(k_i+2)$, so $\sum_i 1/(k_i+2)=1$.  Let $K=\lcm (k_i+2)$ and
$w_i=K/(k_i+2)$; this corresponds to a Fermat hypersurface in
$WP(w_i)$ at a special point in K\"ahler moduli space with $Z_K$
quantum symmetry.

A useful subset of boundary states are the rational $L=0$ states
with various $M$ values.  Their intersection form (counting massless
Ramond strings) is
\eqn\intmatrix{
I = \prod_j (1- g^{w_j}).
}
The field corresponding to $g^{w_j}$ in this
expression (call it $X^i$)
has an NS partner which is non-trivial only in the $j$'th
minimal model, where it is the chiral primary $\phi_{(k_i,k_i,0)}$.
Thus it has $U(1)$ charge $1-2/(k_i+2)=1-2w_i/K$ and corresponds to
a morphism of this degree.  The field $Y_{[ij]}$
corresponding to $g^{w_i+w_j}$
is non-trivial in the other three minimal model factors, and has
$U(1)$ charge $3-2 \sum_{l\ne i,j} w_l/K$.

As argued in \dd\ the leading term in the superpotential is
the cubic term \qcsuper.
The constraints on the $k_i$ guarantee that the degrees of 
$X^i X^j Y_{[ij]}$ add up to $3$ and thus this cubic term
should correspond directly to the flow of a similar term 
computed at large volume.  Perhaps more interestingly,
a product of five fields including one of each $X^i$ also has degree
$3$.  These facts and the direct relation between the gradings of the
morphisms and the $\Delta\grade$ between pairs of branes go a long way
towards guaranteeing that the flow will produce sensible
large volume gradings; however there is still something non-trivial
to check, namely that the winding numbers of the gradings along the
flow are as predicted.

\ifig\quintplot{The evolution of the grades for
fractional branes on the quintic from the Gepner point (where
$\grade(B_k)=2k/5-7/10$) to large volume, where they asymptote to 
$\pm 1/2$.}{\epsfxsize4.0in\epsfbox{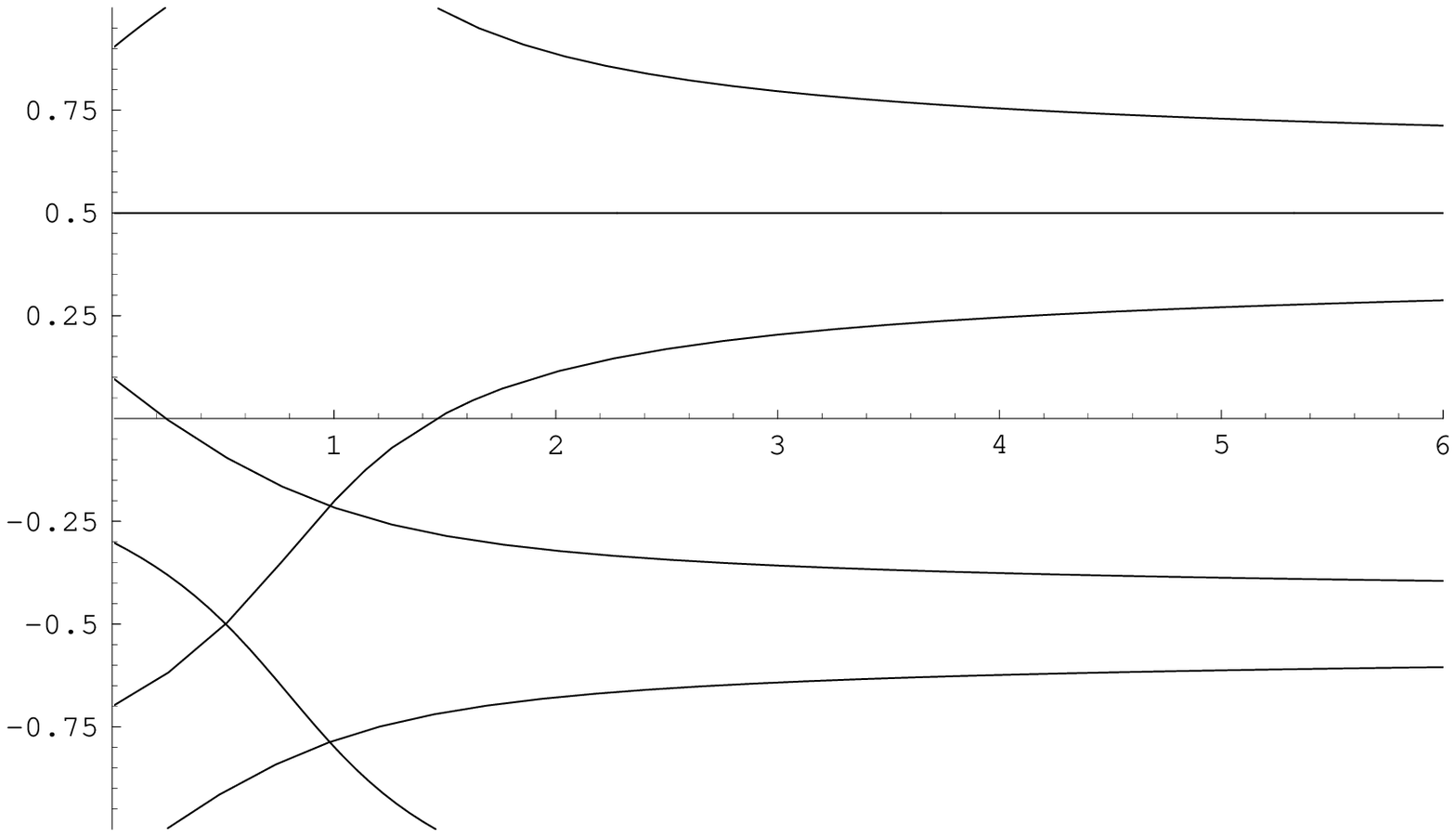}}

We now consider the quintic with $w_i=1$.  The five fractional branes 
are exterior products of a twisted cotangent bundle on $\P^4$,
restricted to the quintic: $B_1=\CO(-1)$, 
$B_2=\bar \Omega^3(3)$,
$B_3=\Omega^2(2)$,
$B_4=\bar \Omega(1)$, and
$B_5=\CO$.
Their grades $\varphi$ 
are plotted as a function of moduli along
the C axis (in the conventions of \dtopics) in \quintplot.

The computation of the morphisms at large volume is straightforward
as the restriction from $\P^4$ is trivial (it is given by tensoring with
the exact sequence 
$$0 \mapr \CO_{\P^4}(-5) \mapr^f \CO_{\P^4} \mapr \CO_X \mapr 0,$$
and for the maps we are considering, the first term gives zero).
By considerations similar to the previous example, we find
$\dim\Ext(B_i,B_{i+1}[-1]) = 5$ for $1\le i\le 4$, and
$\dim\Ext(B_5,B_1[2]) = 5$.
We also have the composition of two successive $\Hom$'s which give
$\dim\Hom(B_i,B_{i+2}) = 10$ and its Serre dual
$\dim\Ext(B_{i+2},B_i[2]) = 10$ for $1\le i\le 3$,
and three $\Hom$'s giving
$\dim\Ext(B_i,B_{i+3}[-1]) = 10$ for $1\le i\le 2$.

Going to the orbifold point, one has $\Delta(\grade_{i+1}-\grade_i)=3/5$
for the pairs $1\le i\le 4$, leading to the first four degree $3/5$
morphisms $X$, while $\Delta(\grade_1-\grade_5)=3/5-3$ (since the sum is
zero) leading to the fifth $X$.  The $Y$ morphisms work similarly
(since the sum of degrees around each $XXY$ triangle is $3$).

\newsec{Physical branes as a subcategory of topological branes}

We now address the question of how to identify the physical branes
at a specific point in K\"ahler moduli space $\MK$.

One approach to this uses the close analogy between general lines
of marginal stability, and the ``wall crossing'' associated with
variation of $\mu$-stability, which determines whether a holomorphic
bundle admits a Hermitian Yang-Mills connection.  This idea was developed
in \pistable\ into the proposal that BPS branes are the $\Pi$-stable objects.
An object $E$ is $\Pi$-stable if every subobject $E'$ (i.e. one for which
there is an injective $\Hom(E',E)$) satisfies $\grade(E')<\grade(E)$.
Further discussion of this idea appears in \fiolmarino.

Although well motivated, the difficulty in combining this proposal with
the present considerations is that
the derived category is not an abelian category and
does not have a notion of ``kernel'' or ``subobject.''
The basic reason for this is extremely simple:
since we identify $A$ with any combination $AB\bar B$ 
where $B\bar B$ is the trivial brane-antibrane
bound state (a complex with the identity map), any brane $B$
appears to be a subobject of $A$.

This point is not just academic, as one can check in examples
(as we will do shortly) that the subobject relation is different at
different points in $\MK$.

To better understand this point, we will need to understand what
universal structure underlies exact sequences and subobjects in the
derived category, and how different abelian categories can sit in the
same derived category.

\subsec{Triangulated categories and bound state formation}

In the references, it is shown that derived categories are not
abelian; in particular there is no idea of exact sequence.  Since
every subobject relation $0 \mapr E' \mapr E$ can be completed to an
exact sequence, this is the key point in making sense of
$\Pi$-stability in this general context.

The closest analog to exact sequences are the ``distinguished triangles.''
Let us explain what these are in the context of the
the derived category $D(\CA)$ formed from the abelian category $\CA$.
(One can axiomatize this structure and talk about triangulated categories
which are not necessarily derived in this way, but we won't use that.)

First, for every morphism $A\mapr^f B$ one has a
distinguished triangle
\eqn\disttri{
\ldots\mapr C_f[-1]\mapr^\psi A\mapr^f B\mapr^\phi C_f\mapr \ldots .
}
Here $C_f$ is the ``cone'' of $f$, the quasi-isomorphism class
of complexes with terms
$A[1] \oplus B$ 
and differential $\left(\matrix{d_A&0\cr f& d_B}\right)$.
This is the basic construction we will try to use to represent the
``brane-antibrane bound state'' ${\bar A} B$ produced by condensing the
tachyon $f$.\footnote*{
There is a technical point which all the references emphasize, namely
that this construction is noncanonical: given a specific chain map
$f$, one is making further choices in writing an explicit complex
representing $C_f$.  All of these choices
are quasi-isomorphic, however, so one can speak of
``the'' cone $C_f$ both as an object in the derived category and as
a physical object.}
The sequence \disttri\ repeats (with a shift of grading) indefinitely to
the left and right and this is why it is better thought of as a triangle.

The expression \disttri\ appears to single out one arrow as special, the
morphism $\psi$ with degree one.  Actually this is only a choice of
notation: for example we could define $D\equiv A[1]$ and find that the
morphism $f$ appeared to have degree one.  If one consistently
identifies shifted branes $B[2n+1]$ with antibranes $\bar B$, the
the relation between the K theory classes of the objects is
independent of this choice.  For example, in \disttri, we have 
$[A] + [C] = [B]$, while
in terms of $BCD$ we would have had $[B] + [D] = [C]$ which
(since $[A]+[D]=0$) is the same.  Similarly $[C[-1]] + [B] = [A]$.

The point now is that if we want to base our discussion on the derived
category, we cannot talk about exact sequences {\it a priori} but must
instead derive them from the distinguished triangles.
If we know that our branes all live in some abelian category within
the derived category, there is a general result which allows us to do
this \bbd: if and only if three successive objects in the triangle all live in
an abelian subcategory $\CA$, that subsequence is a short exact
sequence in $\CA$.  The opposite of this is perhaps easier to see:
the cases which do not correspond to exact sequences are
those in which $C_f$ has homology in both degrees, and thus is not a
member of the abelian category $\CA$.  
This is interesting for our purposes as there will
turn out to be more than one abelian subcategory of $D(\Coh X)$.

Let us discuss this more physically.  Suppose in 
bringing together two objects $\bar A$ and $B$ we can form a bound state $C$,
so that $[\bar A] + [B] = [C]$.  (The choice of notation
$\bar A$ rather than $A$
is just to get the same K theory relation $[B] = [A] + [C]$ we discussed
above.  We are not assuming any specific relation between central charges
yet.)
This must start by condensing a tachyonic or massless
open string between $\bar A$ and $B$.
Let us denote the corresponding morphism (the ``$\Ext$'')
with a double arrow $\Longrightarrow$, while single arrows
denote $\Hom$'s (enhanced gauge symmetries).

From the point of view of BPS central charge, $Z_\bA=-Z_A$ and $Z_B$ might
be aligned, antialigned or not aligned, and the process will look slightly
different depending on this.  If $Z_\bA$ and $Z_B$ are antialigned,
there are two further possibilities $|Z_\bA|>|Z_B|$ or $|Z_\bA|<|Z_B|$, which
determine whether the bound state $Z_C$ aligns with $Z_\bA$ or $Z_B$.
We now explain how all of these possibilities are contained in
the same distinguished triangle \disttri.

If $|Z_B|>|Z_\bA|$, the appropriate description is an exact sequence
\eqn\boundone{
0\mapr A\dmapr^f B\mapr^\phi C\mapr 0
}
with maps $f$ and $\phi$ of degree $0$.
This is the special case in which $C_f$ has homology only in degree $0$,
and every such exact sequence corresponds to a
distinguished triangle \disttri.

If $|Z_\bA|>|Z_B|$, we should use the exact sequence
\eqn\boundtwo{
0\mapr \bar C \mapr^\psi A\dmapr^f B \mapr 0,
}
appropriate if $C_f$ has homology only in degree $-1$.  Here we have
identified $C[-1]$ with the antibrane $\bar C$.

Finally, if $Z_\bA$ and $Z_B$ are aligned, we have
\eqn\boundthree{
0 \mapr B \mapr^\phi C \mapr^\psi \bar A \mapr 0
}
in which $\bar A \dmapr^f B[1]$ does not even appear in the exact sequence
(it is the ``connecting map'' of the long exact sequence).

The point of all this is to show that the derived category
can remain invariant under varying K\"ahler moduli, while describing
somewhat different looking physical processes.  
The distinction between the processes comes when we identify a
specific exact sequence in the triangle.
In all three limiting cases (in which central charges align), 
the exact sequence is the subsequence of \disttri\ containing the maps of
degree $0$, which is the subset of objects which can
live in the same abelian category.

There is some correspondence between our string-inspired
notation and the usual mathematical notation and one might try to
identify our double arrows $\Longrightarrow$ with the special
(degree one) morphisms of \disttri.  Again, one must recognize that this is
only a choice of notation: a single distinguished triangle admits all
three interpretations in which any of the links is a matter field
(this could presumably be proven from crossing symmetry of the
related BPS
algebra \hm).  More importantly, the ``special'' morphism
in our notation need not have degree $1$; this will change under flow.

\subsec{Examples}

Let us now look at some examples where we know the interpretations
on both sides of a flow.

The simplest situation is illustrated by the brane-antibrane pairs
which come out of the Beilinson and generalized McKay constructions.
For example, in $\BC^3/\BZ_3$, we have a tachyon $f\in\Ext^1(B_1,B_2[-1])$
at large volume (where $B_1=\CO(-1)$ and $B_2=\bar\Omega(1)$), which
flows to a massless field $f\in\Ext^1(B_1,B_2)$ at the orbifold point.

At large volume, the BPS central charge is dominated by the D$4$-brane
charge (the rank of the bundle).
Since $\Omega(1)$ is a rank $2$ bundle while $\CO(-1)$ has rank $1$,
at large volume there is a brane-antibrane bound state $X$, 
which is an antibrane (D$4$
charge $-1$) with these conventions.  Its central charge aligns
with that of $B_2$, so the bound state formation is described by
the exact sequence
\eqn\bbarseq{
0 \mapr \bar B_1 \dmapr^f B_2 \mapr^\phi X \mapr 0 .
}
with $\phi$ a degree zero $\Hom$.

In moving to the orbifold limit, the grade of $B_1$ decreases by $1$,
which results in $f$ of degree $1$, $\psi$ of degree $0$ and $\phi$ of
degree $0$.  The bound state is still $X$, but we now want to
interpret it as a brane-brane bound state or extension
\eqn\bbseq{
0 \mapr B_2 \mapr^\phi X \mapr^\psi B_1 \mapr 0
}
with $f$ the connecting map $B_1\dmapr B_2[1]$.

Both come from the distinguished triangle
$$
\ldots \mapr B_1[-1] \dmapr^f B_2 \mapr^\phi X \mapr^\psi B_1 \mapr \ldots 
$$
by specializing to the triple involving maps of degree zero or
equivalently whose terms all have the same grade.  Note that in the
\bbarseq\ interpretation, the object $\bar B_1$ has the same K theory
class as $B_1[-1]$ (this does not change under flow) but grade zero
instead of one.

In the language of abelian categories, the difference is that at
large volume $\bar B_1$ and $B_2$ are both in the abelian category
(justifying the use of \bbarseq) while at the orbifold point $B_1$ and
$B_2$ are in the abelian category (justifying \bbseq).

We will now try to regard this as a valid description along the flow,
where in general the maps will not have integral degree.  Of course
this is what we expect from the CFT discussion, but it means that we
cannot {\it a priori} rely on the exact sequence interpretation of
either \bbarseq\ or \bbseq.  Nevertheless, physics tells us that we
should regard this triangle as describing the formation of a bound
state which exists all along the flow.  Since the flow is continuous,
the grade of $X$ is everywhere determined -- this is not something we
could take for granted without the triangle, as there is no obvious
canonical way to assign gradings to general complexes (they would have
to depend on individual homologies, not the total K theory class).
Since we have the triangle, we can assign $X$ a grade, and one finds that
the gradings of all three morphisms stay within the interval $[0,1]$
all along the flow.  This is consistent with all three objects remaining
stable along the flow (in principle they could still be destabilized by
other objects).

We next discuss the two-brane, the structure sheaf $\CO_\Sigma$ of
a two-cycle $\Sigma$.  There is a simple exact sequence
which produces it from the $B_i$, namely
$$
0 \mapr \CO(-1) \mapr^f \CO \mapr^g \CO_\Sigma \mapr 0 ,
$$
and the corresponding distinguished triangle is
$$
\CO(-1) \mapr^f \CO \mapr^g \CO_\Sigma \mapr^h \CO(-1)[1] \mapr \ldots .
$$
At large volume this is a perfectly good representation of the 2B;
the definition \defgrade\ leads us (as discussed in section 4) to
assign it grade $1/2$ (in conventions where $\grade(\CO)=0$) and thus
the maps $g$ and $h$ have degree $1/2$.  Thus at and near large volume,
the 2B can be produced as a bound state of 4B's by condensing a tachyonic
open string, as discussed in \oz.

As we flow down, the grade of $\CO(-1)$ decreases and the grade of
$\CO$ increases.  Eventually these reach $-1/2$ and $1/2$
respectively, while the 2B remains at $1/2$.  At this point, $f$ has
degree $1$, and $g$ and $h$ have degree $0$.  If we pass this point,
the degree of $g$ goes negative and the 2B goes unstable, while the
open string between 4B's becomes massive.

This conclusion basically agrees with that of \noncompact, where it
was justified in terms of $\Pi$-stability.  However there was an
interesting subtlety noted there; the map $g$ does not look injective
when one follows the usual large volume definitions (a map from a sheaf
to a sheaf with lower dimensional support could hardly be injective).
In the present discussion, since both $g$ and $h$ have degree $0$
at the transition, one wants to interpret them as forming an exact
sequence at that point, in which case $g$ would ``become'' an injective
map.  This type of argument can be tested against other known
subobject relations at the orbifold point, and seems to work.
It suggests a reformulation of $\Pi$-stability which we will make below.

By the time we reach the orbifold
point, the map $f$ has degree $2$ and there is no obvious sign that
the 2B ever existed.  However, since we can determine that $f$ existed
from orbifold considerations (it was an $\Ext^2$ in the quiver-complex
formalism), we can run this backwards.  From this point of view, we
would define $\CO_\Sigma$ as a cone $C_f$ in \disttri.
Coming back up, the degree of
$f$ eventually drops to $1$, and it becomes consistent to postulate
that this $\CO_\Sigma$ becomes stable, with maps $g$ and $h$ of
degree $0$.  Thus we could infer the existence of the 2B elsewhere in
moduli space just from information obtained at the orbifold point.

A very similar discussion can be made for the ``mysterious'' bound
state $\ket{10000}$
discussed in \refs{\dtopics,\denev}, which exists at the Gepner
point in the quintic but not at large volume.  This is a bound state
of $\CO(-1)$ and $\CO$ and comes about because an $\Ext^3(\CO,\CO(-1))$
at large volume eventually becomes tachyonic near the orbifold point.
In this example, one can infer the existence of a bound state which is
not a sheaf, just having information about sheaves and about the BPS
central charges.

These results demonstrate how we can infer the existence of new BPS
branes at distant points in $\MK$.  Given a set of BPS branes, we
construct their derived category, including new objects which are
candidate BPS branes elsewhere in moduli space.  The K theory class of
each such object is determined, and this determines the grade of the
connecting maps up to an overall $2\BZ$ ambiguity.  If this ambiguity
can be fixed in a way that gives all the morphisms
non-negative grade, then the object becomes stable.

\subsec{t-structures}

There is a mathematical formalism which makes it possible to identify
abelian categories within the derived category, the formalism of
t-structures \refs{\bbd,\gm}.  We will not actually use this in the
proposal we are about to make, but it illustrates in a clear way how
objects which look like complexes from one point of view can be
individual objects (not complexes) in a different abelian category.

A t-structure is defined by prescribing two subcategories
$\CD^{\ge 0}$ and $\CD^{\le 0}$ of the derived category $\CD$.
If $\CD$ is derived from an abelian category $\CA$, these would
be the categories of complexes with cohomology only in degrees $n\ge 0$
and complexes with cohomology only in degrees $n\le 0$, respectively.
The original abelian category would then be
$\CA = \CD^{\ge 0} \cap \CD^{\le 0}$.

The non-trivial fact is now that there are a short list of axioms which
$\CD^{\ge 0}$ and $\CD^{\le 0}$ must satisfy in order to guarantee that
their intersection is an abelian category.  This provides a general
way to define new abelian subcategories.  In general, the intersection
will contain objects which are complexes from the point of view of the
original category; thus these complexes will be regarded as objects
in the new category.  In this sense, the distinction between ``object''
and ``complex'' is not invariant, and from this point of view it is
quite natural to get complexes of coherent sheaves as stable BPS branes
at other points in $\MK$.

A simple way to get new t-structures on $D(\Coh X)$ for $X$ Calabi-Yau
is to use the Fourier-Mukai transforms discussed earlier.  Taking the
images of $\CD^{\ge 0}$ and $\CD^{\le 0}$ under the FMT provides a new
t-structure whose abelian category, although isomorphic to the original
(for an FMT implementing a monodromy), consists of different objects in
$D(\Coh X)$.  Since these monodromy groups are braid groups this
presumably leads to infinitely many t-structures on $D(\Coh X)$.

This formalism might be directly usable to produce the abelian categories
which are needed by the original formulation of $\Pi$-stability.
The basic way this could work would be to move objects between
$\CD^{\ge 0}$ and $\CD^{\le 0}$ when their gradings flow across zero.

Indeed, a very concrete example illustrating this idea can be found in
work of Bridgeland \bridgeland.  In \bridgeland\ it is proven that
$D(\Coh X)$ is invariant under a flop transition, a previously known
result \boflop, but the interesting point is the way in which this is
proven.  This uses the theory of so-called perverse sheaves, which can
be formulated using t-structures.  The basic idea is to consider some
submanifold (more generally a stratification of the manifold) and
shift the grading of all objects supported on that submanifold in
defining the t-structure.  The corresponding abelian category is
referred to as a category of ``perverse sheaves.''

In \bridgeland, one considers perverse sheaves defined by shifting the
grading of sheaves supported on an appropriate curve $C$ in the CY by
$-1$.  Certain of these perverse sheaves can be identified as
analogous to points, and it turns out that the moduli space of these
``perverse points'' is isomorphic to the CY $X'$ which is the result
of a flop transition on the curve $C$.

The point of contact with our present considerations is that in terms
of periods and BPS central charges, a flop transition simply acts by
taking the central charge $Z=B+iV$ of the 2B on the curve $C$ from
$V>0$ to $V<0$.  This indeed shifts the grading of 2B's by $\pm 1$ 
(the sign depends on the first Chern class) and by arguments
similar to Bridgeland's, might lead to the t-structure appropriate for
the flopped CY.

It seems very likely to us that these ideas will be important in
future work and more specifically that subcategories of $D(\Coh X)$
generated by the stable branes of a given grade on a CY with given
moduli will provide a significant generalization of categories of
``perverse sheaves.''  However, we will leave the problem of making
this more explicit to future work.

\subsec{A proposal for a stability condition}

The stability condition we will propose here builds more simply on the
results we already presented and adds some plausible physical input.
Unlike $\Pi$-stability, it does not provide a way to decide whether a
brane is stable at a point in $\MK$ by just considering that point,
but only describes the variation in the set of stable branes as one
moves in $\MK$.

Many of the previous considerations can be summarized by defining a
``stable triangle.''  A distinguished triangle \disttri\ involves
three morphisms with grades $\alpha+\beta+\gamma=1$.  Let
a stable triangle (at a given point in $\MK$) be a distinguished triangle
for which each of the three grades is in $[0,1]$.

Since if two brane central charges are colinear, the third will be as
well, the definitions lead to the constraint that the only stable triangles
on the boundary of this region (or ``semistable'' triangles) are those
with one grade $1$ and the other two $0$.

We cannot directly use this to say that a stable object only
participates in stable triangles -- there will always be lots of extra
triangles involving negative morphisms.  We only know that every
morphism between stable objects must have non-negative degree.
Indeed, there is no physical argument that there should be
a canonical definition of grade for unstable objects, and we will
not assume that there is.

Let us now suppose that we know the stable objects at some point in
$\MK$, and we move a small distance in this space.  The gradings of
morphisms will flow: some triangles will become unstable and we must
lose objects; others will become stable and we can gain objects.

When a triangle goes unstable, the brane which decays will always be
the one sitting between the two morphisms of zero degree.  This is
because (in all three cases \boundone,\boundtwo, and \boundthree)
this will always be the heaviest of the three branes.
This also means that one only need check lighter branes as possible
destabilizing subobjects.

Conversely, when a triangle becomes stable, we will see it by a
map $f$ between stable objects having degree coming down through $1$.
Whenever this happens, we can try to add $C_f$ with grade which gives the
other maps degree $0$.  This uniquely determines the $2\BZ$ ambiguity
in the grading of $C_f$.  We should however only add $C_f$ as a stable
object if it is not also destabilized by a morphism of negative
degree from some preexisting stable object.  Once we do this, we might
have further candidate bound states involving $C_f$, so the process
must be iterated.

This more or less restates the phenomena we observed in our simple
examples but now we must face the question of whether this procedure
leads to an unambiguous modified list of stable objects or whether the
result depends on the order in which we make these modifications.

One point where such dependence might enter is that we might find
that $A$ destabilizes $B$, but $A$ also decays on the same line.
The general result which prevents this type of ambiguity is that
the subobject of $A$ responsible for the decay will also be a subobject
of $B$ (by composing the $\Hom$'s), and typically would be a stable
object which will destabilize $B$.  In general it might decay, but this
chain must terminate with some stable final product (assuming the
spectrum of masses has a gap) which will also be a subobject.

Similarly, we cannot find that adding a new object $C_f$ destabilizes
preexisting objects, because there will be a subobject of $C_f$ which
already destabilized them before we added $C_f$.

These considerations suggest that the procedure as we stated it is
unambiguous.  The main physical assumption we needed was that the
spectrum of BPS masses has a finite mass gap, so that we cannot have
infinite chains of subobjects and decays.

\newsec{Conclusions}

In this work, we gave a fundamental picture of BPS D-branes on Calabi-Yau
manifolds, based on considerations in conformal field theory and the
related topological string theory.

To summarize, we distinguished boundary conditions in topological open
string theory or ``topological D-branes'' from BPS boundary conditions
in CFT or ``physical D-branes,'' and argued that every physical
D-brane had a topological analog but not vice versa.  Topological
D-branes (in the B model) are more general than holomorphic bundles or
coherent sheaves; they can be arbitrary objects of the derived
category.  The grading of morphisms between topological D-branes
depends on the K\"ahler moduli in a simple way
and this is responsible for variation
of the spectrum of physical D-branes and lines of marginal stability;
branes involved in morphisms of negative degree cannot exist.
This provides a CFT derivation of the $\Pi$-stability condition of
\pistable.  We went on to discuss the
triangulated structure of the derived category, which allows us to
dispense with the requirement of a preexisting abelian category and
subobject relation made in the original $\Pi$-stability proposal,
instead deriving the subobject relations from the gradings and
distinguished triangles.  All of these points were illustrated in a
number of examples; in simple cases these ideas lead directly to
explicit predictions for marginal stability lines.  

Many new phenomena are clearly possible and can now be studied
systematically, such as the formation of branes (away from the large
volume limit) which are not coherent sheaves but more general objects
in the derived category.

All of these developments appear rather solid to us and provide a firm
basis for further understanding of BPS branes on Calabi-Yau as well as
a precise contact with the homological mirror symmetry proposal of
Kontsevich.  The ``flow of gradings'' is a new structure in this problem
which we believe will be quite important in future developments.

In the final subsection, we went further and stated a definite
proposal for how to determine the spectrum of BPS branes at arbitrary
points in K\"ahler moduli space.  This proposal is somewhat harder to
use than $\Pi$-stability in that it requires starting with the
spectrum at a single point, say the large volume limit, and following
its evolution to the point of interest.  (It is not necessary to
follow the entire spectrum in order to determine the existence of
particular branes, however.)  We did not prove that this
procedure always leads to unambiguous results, though we did give
suggestive arguments for this.

Clearly this proposal requires a great deal of testing and exploration
at this point.  There are numerous self-consistency checks that it
must pass; for example it is not obvious that branes whose periods
vanish at non-singular points of K\"ahler moduli space will decay
before reaching these points (as is required for physical
consistency).  We did not even prove that monodromies are symmetries
of the physical spectrum.

Not having yet performed these basic checks, our main reason for
believing in the proposal at present is that it seems to us to be the
conceptually simplest proposal which could accomodate the
known complexity of these problems as revealed in \noncompact\ and our
further studies.  Since it is the first such proposal, this point will
have to be confirmed by further work as well.  Hopefully there is a lot
of scope for simplifying its application; ideally one would be able
to derive a condition which can be applied at a single point in
K\"ahler moduli space.  One might well benefit from using more A model
information as well.

We will not get into lengthy discussion of the likely applications of
this work here, instead referring to the conclusions of \noncompact.
Perhaps the most direct application would be to provide a simpler
invariant of $d=4$, $\CN=2$ string compactifications than the explicit
spectrum of BPS branes, namely the derived category obtained from this
spectrum.  The precise sense in which this is simpler is that it does
not depend on the BPS central charges or the point in vector multiplet
moduli space.  Making interesting use of this idea in studying $\CN=2$
duality probably requires generalizing the ideas to defining ``derived
categories of quantum BPS branes,'' which would have some similarity
to BPS algebras \hm\ but presumably would be independent of vector
moduli.

As discussed in the conclusions to \noncompact, we regard the more
important goal of this line of work to be its eventual application to
understanding $\CN=1$ compactifications of string theory.  Building on
\pistable, in $\CN=1$ language we have provided a rather complete
discussion of the problem of solving the D-flatness conditions in a
certain large class of theories.  As will be discussed elsewhere, we
believe it will turn out to be possible to get exact superpotentials
in Gepner models and perhaps more general CY's as well.  

A natural next question in this vein is whether a similar geometric
understanding could be developed of non-BPS branes.  One should
distinguish two cases.  The examples we know of are connected to BPS
branes or combinations of BPS branes by varying CY moduli, and it
seems very likely that these can be understood in the same way, with
the non-BPS property arising from spontaneous breaking of space-time
$\CN=1$ supersymmetry, now involving a competition between D and F
flatness conditions.  There might be other non-BPS branes not
connected to BPS branes by varying moduli; for these it is unclear
whether such a picture would apply.

In any case, we believe our present results give further evidence that
$\CN=1$ string compactification can lead to problems which admit
general solutions (not just case by case analysis) and a rich
mathematical structure.

\medskip

We would like to thank T. Banks, N. Berkovits, F. Bogomolov, T. Bridgeland,
I. Brunner, P. Deligne, D.-E. Diaconescu, B. Fiol, A. Kapustin,
S. Katz, A. Klemm, M. Kontsevich, W. Lerche, C. L\"utken, M. Mari\~no,
P. Mayr, G. Moore, D. Morrison, Y. Oz, A. Polishchuk, C. R\"omelsberger,
A. Sen, P. Seidel, E. Sharpe, R. Thomas, A. Tomasiello, and
B. Zwiebach for helpful discussions and comments throughout the course
of this work.  I particularly thank R. Thomas for critical remarks which
are addressed in the revised version.

This research was supported in part by DOE grant DE-FG02-96ER40959,
and by the Clay Mathematics Institute.

\listrefs
\end